\documentclass[a4paper,11pt]{article}
\pdfoutput=1
\usepackage{jheppub}
\usepackage[T1]{fontenc}
\usepackage{lmodern}
\usepackage{cleveref}
\crefname{figure}{figure}{figures}
\crefname{appendix}{appendix}{appendices}
\crefname{subappendix}{appendix}{appendices}
\usepackage{array}
\usepackage{calc}
\usepackage{longtable}
\usepackage{booktabs}
\usepackage{multirow}
\usepackage{slashed}
\usepackage{xcolor}
\usepackage{graphicx}
\usepackage{float}
\usepackage{placeins}
\usepackage{xspace}
\usepackage{units}
\usepackage[format=hang,labelfont=bf,hypcap=true]{caption}
\usepackage{subcaption}
\usepackage{enumitem}
\AddToHook{bfseries}{\boldmath}

\allowdisplaybreaks[2]
\newcommand{\ie}{\textit{i.e.\@}\xspace}

\newcommand{\Vegas}{V\protect\scalebox{0.8}{EGAS}\xspace}

\newcommand{\KKMC}{K\protect\scalebox{0.8}{KMC}\xspace}
\newcommand{\BabaYaga}{B\protect\scalebox{0.8}{ABA}Y\protect\scalebox{0.8}{AGA}\xspace}
\newcommand{\McMule}{M\protect\scalebox{0.8}{C}M\protect\scalebox{0.8}{ULE}\xspace}
\newcommand{\Phokhara}{P\protect\scalebox{0.8}{HOKHARA}\xspace}
\newcommand{\eik}{\mathfrak{s}}
\newcommand{\nmax}{n_{\max}}
\newcommand{\Ninja}{N\protect\scalebox{0.8}{INJA}\xspace}
\newcommand{\GoSam}{G\protect\scalebox{0.8}{O}S\protect\scalebox{0.8}{AM}\xspace}
\newcommand{\Golem}{G\protect\scalebox{0.8}{OLEM}95\xspace}

\newcommand{\Collier}{C\protect\scalebox{0.8}{OLLIER}\xspace}

\newcommand{\FORM}{F\protect\scalebox{0.8}{ORM}\xspace}

\newcommand{\Sherpa}{S\protect\scalebox{0.8}{HERPA}\xspace}

\long\def\symbolfootnote[#1]#2{\begingroup%
\def\thefootnote{\fnsymbol{footnote}}\footnote[#1]{#2}\endgroup}

\newcommand{\abs}[1]{\left| #1\right|}
\newcommand{\rbr}[1]{\left( #1\right)}

\newcommand{\cbr}[1]{\left\{ #1\right\}}
\newcommand{\sbr}[1]{\left[ #1\right]}

\makeatletter
\providecommand*{\diff}%
{\@ifnextchar^{\DIfF}{\DIfF^{}}}
\def\DIfF^#1{%
\mathop{\mathrm{\mathstrut d}}%
\nolimits^{#1}\gobblespace}
\def\gobblespace{%
\futurelet\diffarg\opspace}
\def\opspace{%
\let\DiffSpace\!%
\ifx\diffarg(%
\let\DiffSpace\relax
\else
\ifx\diffarg[%
\let\DiffSpace\relax
\else
\ifx\diffarg\{%
\let\DiffSpace\relax
\fi\fi\fi\DiffSpace}
\makeatother

\providecommand{\xmin}{\ensuremath{x_{\text{min}}}}
\providecommand{\xmax}{\ensuremath{x_{\text{max}}}}

\newcommand{\ptilde}[1]{{\tilde{p}_{#1}}}

\newcommand{\done}{{\rm d}}

\newcommand{\bea}{\begin{eqnarray}}
\newcommand{\eea}{\end{eqnarray}}
\newcommand{\bi}{\begin{itemize}}
\newcommand{\ei}{\end{itemize}}

\newcommand{\DiLog}[1]{\text{Li}_2\left(#1\right)}

\newcommand*{\TeV}{\ensuremath{\text{Te\kern -0.1em V}}}
\newcommand*{\GeV}{\ensuremath{\text{Ge\kern -0.1em V}}}
\newcommand*{\MeV}{\ensuremath{\text{Me\kern -0.1em V}}}
\newcommand*{\keV}{\ensuremath{\text{ke\kern -0.1em V}}}
\newcommand*{\eV}{\ensuremath{\text{e\kern -0.1em V}}}

\newcommand{\LLs}{LL$_\text{s}$\xspace}
\newcommand{\NLLs}{NLL$_\text{s}$\xspace}
\DeclareMathOperator{\artanh}{artanh}

\newlist{myitemize}{itemize}{3}
\setlist[myitemize]{leftmargin=14em}

\newcolumntype{C}{>{\centering\arraybackslash}p{0.14\textwidth}}

\newlength{\unitcharwidth}
\usepackage{tikz}
\usetikzlibrary{decorations.pathmorphing,decorations.markings,arrows.meta,calc,positioning,fit,shapes.misc,matrix,backgrounds}

\usepackage{etoolbox}
\makeatletter
\AtBeginDocument{%
  \def\NAT@spacechar{~}%
  \let\NAT@space\NAT@spacechar
  \def\NAT@penalty{\penalty\@M}%
  \patchcmd{\NAT@citexnum}{--\NAT@penalty}{\mbox{--}}{\typeout{natbib range patch OK}}{\typeout{natbib range patch FAILED}}%
}
\makeatother
\definecolor{linkblue}{HTML}{1F3A7A}
\usepackage[colorlinks=true]{hyperref}
\hypersetup{
colorlinks=true,
linkcolor=linkblue,
linktocpage=true,
citecolor=violet,
urlcolor=linkblue
}

\newcommand{\liverpool}{Department of Mathematical Sciences, University of Liverpool, \\
Liverpool L69 3BX, U.K.
}
\newcommand{\salamanca}{Departamento de Física Fundamental and IUFFyM, Universidad de Salamanca, \\
Salamanca E-37008, Spain}

\preprint{MCnet-26-37}
\title{Soft-photon resummation for low-energy $e^+e^-$ colliders}
\author[a]{Thomas Dave,}
\author[a]{Lois Flower,}
\author[a]{J\'er\'emy Paltrinieri}
\author[a,b]{and William J. Torres~Bobadilla}
\affiliation[a]{\liverpool}
\affiliation[b]{\salamanca}
\emailAdd{t.dave@liverpool.ac.uk}
\emailAdd{lois.flower@liverpool.ac.uk}
\emailAdd{jeremy.paltrinieri@liverpool.ac.uk}
\emailAdd{william.torres@usal.es}

\abstract{
We present resummed predictions for $e^+e^- \rightarrow \mu^+ \mu^-(\gamma)$ and $\pi^+ \pi^-(\gamma)$ in the multi-photon soft limit, matched to exact NLO. The resummation is performed within the Yennie--Frautschi--Suura framework at the amplitude level. The initial and final-state contributions are included coherently, and NLO accuracy with finite electron, muon and pion masses is achieved. We present our calculation, its implementation in a fast Fortran Monte Carlo integrator, and compare our predictions with the generators \BabaYaga, \KKMC, \Phokhara and \McMule, in scenarios inspired by the energy-scan (CMD) and radiative-return (KLOE, BES\,III, B factories) experiments. We find agreement with \BabaYaga, which includes higher-order corrections through a parton shower, at the per-mille level in the cross-sections, and differences of up to two percent with the soft-photon resummation of \KKMC. This work represents an important step towards the validation of Monte Carlo generators for low-energy experiments, and will be released in the upcoming new version of \Phokhara.
}

\begin{document}
\maketitle
\flushbottom
%= text ===========================================
\section{Introduction}
\label{sec:introduction}

Modern particle physics has decisively entered a precision era, in which previously neglected subleading contributions to theory calculations must be taken into account to match experimental precision. In addition, increased experimental precision has worsened a number of tensions, most notably the hadronic contribution to the muon anomalous magnetic moment, $a_\mu^\text{HVP}$ \cite{Aliberti:2025beg}. While lattice QCD predictions agree with the measured experimental value, unresolved tensions within the data-driven extraction of $a_\mu^\text{HVP}$ still exist and render the status of the muon anomalous magnetic moment in the Standard Model unclear. These tensions originate from discrepancies among the measurements of the $e^+e^-\to\text{hadrons}$ cross-section~\cite{CMD-3:2023alj,KLOE-2:2017fda,BaBar:2012bdw,BESIII:2015equ,SND:2020nwa}, most notably in the $\pi^+\pi^-$ channel, which accounts for the bulk of $a_\mu^\text{HVP}$ and of its uncertainty. Many of these measurements predate the recent progress in theory calculations and simulation, and efforts are underway to revisit the experimental analyses with up-to-date Monte Carlo tools containing state-of-the-art radiative corrections. These tools are part of the measurement itself, since their accuracy enters directly into the extracted cross-sections and, in particular, their systematic uncertainty: unless higher perturbative orders, the resummation of large logarithms and improved models of non-perturbative effects are all included in them, theory, rather than experiment, will set the limit on the precision of the hadronic cross-sections~\cite{Aliberti:2024fpq}. The same cross-sections fix the hadronic contribution to the running of the electromagnetic coupling, $\Delta\alpha_{\rm had}(M_Z^2)$, an input of the electroweak precision tests~\cite{WorkingGrouponRadiativeCorrections:2010bjp,Keshavarzi:2019abf,Davier:2019can} whose uncertainty will limit the FCC-ee programme~\cite{Proceedings:2019vxr,FCC:2025lpp,Placzek:2026hvx}; the inclusion of precise radiative corrections serves two precision programmes.

Two paradigms are used to extract the dependence of the hadronic cross-section on the effective centre-of-mass energy: the energy scan, where the collider energy is varied, and the radiative return~\cite{Binner:1999bt,Benayoun:1999hm}, where the collider energy is fixed and the emission of hard initial-state radiation lowers the invariant mass of the final-state charged pair.
In the latter, the radiative process is thus the lowest-order contribution.
A number of Monte Carlo tools describe the processes $e^+e^-\to\mu^+\mu^-(\gamma)$ and $\pi^+\pi^-(\gamma)$ at low energies in one or both of these modes.
\BabaYaga~\cite{Balossini:2006wc,Budassi:2026lmr}, developed originally for luminometry processes, calculates lepton and pion pair production in both the scan and radiative-return modes at NLO matched to a QED parton shower. \KKMC~\cite{Jadach:1999vf,Jadach:2022mbe} has long been the reference for lepton pair production at energies above 2~GeV, with the NLO corrections matched to the Yennie--Frautschi--Suura (YFS) exponentiation in the scan mode, though only partially included in the radiative return. Its precision at lower energies is limited, since vacuum polarisation (VP) effects are not included. \Phokhara~\cite{Rodrigo:2001kf,Czyz:2003ue,Campanario:2013uea}, the most widely used tool for pion production at radiative-return experiments, implements $e^+e^-\to\mu^+\mu^-\gamma$ and $\pi^+\pi^-\gamma$ to exact NLO. \McMule~\cite{Banerjee:2020rww,Broggio:2022htr,Ulrich:2025fij} provides fixed-order corrections at NNLO for the scan processes and at NLO for the radiative return, complete for the muons and restricted to the initial state for the pions.
As this list shows, the radiative return is limited to fixed-order NLO accuracy in most of these tools, and the way towards full NNLO is still in the making. \McMule has recently provided the NNLO initial-state corrections (ISC) to $e^+e^-\to\mu^+\mu^-\gamma$~\cite{Fang:2026dbs}, in which the two-loop amplitudes, known only for a massless electron~\cite{Fadin:2023phc,Badger:2023xtl}, are supplemented with the electron-mass logarithms through massification~\cite{Penin:2005eh,Becher:2007cu,Engel:2018fsb}, which restricts the predictions to selections with a photon tagged at large angle. The NNLO ISC under way for \Phokhara instead retain the full dependence on the electron mass, and their first building blocks, the planar two-loop Feynman integrals~\cite{PetitRosas:2025xhm,Pozzoli:2026eiu} and the one-loop amplitudes to higher orders in the dimensional regulator~\cite{Dave:2026pvq}, are already available. This will give access to all experimental scenarios, including those in which a photon is emitted collinear to the beams.

Fixed-order calculations alone, however, cannot reach the required accuracy everywhere in phase-space. The emission of photons with energies $E_\gamma\ll\sqrt s$ generates logarithms of the photon energy $L_s = \ln(\sqrt{s}/E_\gamma)$, each enhanced by the collinear logarithm $L_c=\ln(s/m_e^2)$, which is large because the electron is light, $L_c\simeq14$--$20$ for $\sqrt s=0.7$--$10$~GeV. Every additional photon then brings a factor of order $(2\alpha/\pi)(L_c-1)L_s$ in the cross-section, which becomes large wherever the selection cuts restrict the energy of the radiation, as in the energy scan or near the kinematic edges of the distributions. In the soft limit the amplitudes factorise, and these logarithms can be resummed to all orders by exponentiation, as shown by Yennie, Frautschi and Suura~\cite{Yennie:1961ad}. We refer to the terms of order $\alpha^nL_s^n$, with full mass dependence, as the leading (soft) logarithms (\LLs).\footnote{We count the logarithmic accuracy in the soft logarithm $L_s$. In part of the YFS literature, LL and NLL refer instead to collinear approximations of the higher-order corrections.} When matched to exact NLO in the YFS scheme, the terms of order $\alpha^nL_s^{n-1}$, which we call \NLLs, are also resummed by construction. Their resummation reproduces the NNLO corrections, and those of all higher orders, in the limit in which the additional photons are soft. What it misses at NNLO are the contributions without soft enhancement, such as the two-loop amplitudes beyond their infrared (IR) poles and the one-loop amplitudes with an additional hard photon. These are precisely the ingredients provided by the fixed-order NNLO calculation, which can be systematically included in the soft-photon resummation approach, raising its accuracy to NNLO+N\NLLs by the same mechanism. The collinear logarithm $L_c$ itself is not resummed in this approach: it enters through the exact mass dependence of the eikonal factors and of the NLO matrix elements, so that the hard collinear terms without soft enhancement, $\alpha^2L_c^2$ and beyond, are not captured.

Several codes already resum the radiative corrections beyond fixed order: \BabaYaga \cite{Balossini:2006wc,Budassi:2026lmr} matches its NLO calculation to a QED parton shower based on collinear factorisation, while \KKMC and \Sherpa~\cite{Sherpa:2024mfk,Krauss:2022ajk} exponentiate the soft-photon divergences in the YFS scheme, recently at the amplitude level and extended to the pion channel~\cite{Delorme:2026qbm,Price:2026qop}. The RadioMonteCarLow~2 comparison of ref.~\cite{Aliberti:2024fpq} confronted these generators with common inputs and cuts in the scenarios of the experiments, and found agreement at the per-mille level in the cross-sections but differences of several percent in the distributions sensitive to higher orders.

\bigskip%
In this work we present a new implementation of the coherent exclusive exponentiation (CEEX)~\cite{Jadach:2000ir} for the four processes $e^+e^-\to\mu^+\mu^-(\gamma)$ and $\pi^+\pi^-(\gamma)$ at NLO+\NLLs. Since CEEX exponentiates the soft photons at the amplitude level, the interference between the ISC and the final-state corrections (FSC) is kept at all orders. All the ingredients are computed in-house in dimensional regularisation. The one-loop amplitudes keep the full mass dependence; they are obtained with a four-dimensional tensor decomposition or diagram by diagram, cross-checked against \GoSam~\cite{GoSam:2014iqq}, and built on a single spinor basis, so that the CEEX residuals, calculated with mapped momenta~\cite{Price:2025fiu} that preserve the soft and collinear limits, are free of spurious phases. The YFS form factor, which collects the IR singularities of the real and virtual soft-photon contributions, is computed with the same regulator, so that the IR poles cancel at every phase-space point, and the exact two-photon matrix element improves the description of hard real radiation beyond the eikonal approximation. These ingredients are combined in a fast Fortran Monte Carlo integrator, with a multi-channel phase-space generator~\cite{Kleiss:1994qy} and dedicated analysis routines. We validate the implementation through the cancellation of the IR and UV poles, the independence of predictions on the soft-photon cut, and the comparison of the YFS form factor with the literature. We then compare the predictions of this method with those of \BabaYaga, \KKMC, \Phokhara and \McMule. The implementation will be part of the upcoming release of \Phokhara, which will then provide both this resummation and the NNLO ISC on top of its NLO matrix elements.

The paper is structured as follows: in \cref{sec:methods}, we describe the formalism of the CEEX soft-photon resummation and its implementation in a Monte Carlo integrator, followed by consistency checks of the code in \cref{sec:low-level-validation}. \Cref{sec:results} contains our main results and comparisons with other generators in the scan mode of CMD and in the radiative return of KLOE, BES\,III and the B factories. Finally, we conclude in \cref{sec:conclusions}.
%\section{Formalism and implementation}
\section{Coherent exclusive exponentiation}
\label{sec:methods}

To keep the paper self-contained, this section describes the formalism of CEEX~\cite{Jadach:2000ir} as implemented in
this work. The YFS theorem factorises the IR singularities of the
QED amplitudes to all orders and exponentiates them, leaving IR-finite remainders that can be calculated order by
order. The CEEX formalism extends the YFS theorem by implementing
the factorisation at \emph{amplitude level}, which retains initial--final
interference terms at all orders. We consider the scattering processes
\begin{equation}
  e^-(p_1)+e^+(p_2)\;\to\;f(p_3)+\bar f(p_4)+\gamma(k_1)+\cdots+\gamma(k_n)\,,\qquad
  f=\mu,\pi\,,\quad n=0,1,2,\dots\,,
  \label{eq:process-n}
\end{equation}
with massive charged particles and massless photons,
$p_1^2=p_2^2=m_e^2$, $p_3^2=p_4^2=m_f^2$ and $k_i^2=0$, and centre-of-mass energy squared
$s=(p_1+p_2)^2$. We denote by $\mathcal M_n$ the amplitude of the process with $n$ photons in
\cref{eq:process-n}, ultraviolet (UV) renormalised and with its virtual IR divergences removed, so that
$\mathcal M_n$ is UV and IR finite (see \cref{subsec:amps,subsec:irsub}).
Depending on the experimental setup (see \cref{sec:low-level-validation,sec:results}),
the lowest-order process may be defined either as the $2\to 2$ process without photon emission,
usually referred to as the \emph{energy-scan} process, or as the $2\to 3$ process with the emission of one
\emph{hard} photon, referred to as the \emph{radiative-return} process.
In both cases, additional photons can be included according to the exact matrix elements up to the matching
order, and using the eikonal approximation thereafter, so that an arbitrary
number of photons is generated on top of the Born process.

In the following, we present the theoretical formalism (\cref{subsec:resummation}) together
with the ingredients needed to perform the soft-photon resummation: the real and virtual amplitudes for the
processes in \cref{eq:process-n}, including their IR subtraction and renormalisation
(\cref{subsec:amps,subsec:amplitudes}), the momentum mappings used to include additional photons
and to generate the multi-photon phase-space (\cref{subsec:mappings}), and finally our in-house
implementation of the formalism (\cref{subsec:implementation}).
%a brief discussion of the software implementation
%is included later in this section.
% \Cref{subsec:resummation} presents the
% master formula and the residuals, \cref{subsec:yfs} the YFS form factor,
% \cref{subsec:mappings} the momentum mappings on which the residuals are evaluated,
% \cref{subsec:amplitudes} the one-loop amplitudes, \cref{subsec:irsub} their
% renormalisation and IR subtraction, \cref{sec:ps} the generation of the phase
% space and \cref{subsec:implementation} the structure of the code.

%\subsection{Master formula}
\subsection{Formalism \& Master formula}
\label{subsec:resummation}

Consider the emission of a photon with momentum $k$ and helicity $\sigma$ off the external
charged legs of the processes in \cref{eq:process-n}. In the soft limit, $k\to0$, the amplitude with
$n+1$ photons factorises into the amplitude with $n$ photons times a universal eikonal factor
\begin{equation}
  \mathcal M_{n+1}(p;k_1,\dots,k_n,k)\;\xrightarrow[k\to0]{}\;\eik_\sigma(k)\,\mathcal M_n(p;k_1,\dots,k_n)\,,
  \qquad \eik_\sigma(k)=\eik^{\rm I}_\sigma(k)+\eik^{\rm F}_\sigma(k)\,,
  \label{eq:soft-limit}
\end{equation}
where $p$ collectively denotes $p_1,\dots,p_4$, and $\eik^{\rm I}$ and
$\eik^{\rm F}$ are the amplitude-level eikonal factors for emission from initial and
final states, respectively.\footnote{This factorisation holds both for the amplitudes $\mathcal M_n$ and for the renormalised
amplitudes before their virtual IR divergences are removed through the factor $e^{-\alpha B}$
(see \cref{subsec:irsub}), since the virtual form
factor $B$, defined below in \cref{eq:BV}, depends only on the momenta of the charged legs.}

For two charged initial legs $p_1,p_2$ and two charged
final legs $p_3, p_4$, the eikonal factors are given by
\begin{equation}
  \eik^{\rm I}_\sigma(k)=-\,eQ_e\sbr{\frac{p_1\cdot\varepsilon^*_\sigma(k)}{p_1\cdot k}
                                  -\frac{p_2\cdot\varepsilon^*_\sigma(k)}{p_2\cdot k}},\qquad
  \eik^{\rm F}_\sigma(k)=+\,eQ_f\sbr{\frac{p_3\cdot\varepsilon^*_\sigma(k)}{p_3\cdot k}
                                  -\frac{p_4\cdot\varepsilon^*_\sigma(k)}{p_4\cdot k}},
  \label{eq:eikonal}
\end{equation}
where $\varepsilon_\sigma(k)$ is the polarisation vector of the photon. The sign of each term is
$\theta_iQ_i$, with $\theta_i=+1$ for an outgoing and $-1$ for an incoming leg and $Q_i$ its charge,
so that the relative sign between $\eik^{\rm I}_\sigma$ and
$\eik^{\rm F}_\sigma$ is physical and fixes the sign of the initial--final interference,
following the conventions of ref.~\cite{Jadach:2000ir}. The squares of these eikonal factors,
\begin{equation}
  \sum\nolimits_{\sigma}\abs{\eik^{\rm I}_\sigma(k)}^2
  =-e^2Q_e^2\Big(\tfrac{p_1}{p_1\cdot k}-\tfrac{p_2}{p_2\cdot k}\Big)^2 ,\quad
  \sum\nolimits_{\sigma}\abs{\eik^{\rm F}_\sigma(k)}^2
  =-e^2Q_f^2\Big(\tfrac{p_3}{p_3\cdot k}-\tfrac{p_4}{p_4\cdot k}\Big)^2 ,
  \label{eq:eikonal-sq}
\end{equation}
are the classical soft-photon spectra of the two dipoles. 

The soft-photon factorisation of \cref{eq:soft-limit}, together with the abelian nature
of QED, which implies the factorisation of multiple soft photons, allows the soft
real and virtual emissions to be summed to all orders. The resulting all-order cross-section
reads
\begin{equation}
  \begin{aligned}
  \sigma={}&\frac{1}{\text{flux}(s)}\sum_{n=0}^{\infty}\frac1{n!}
  \int\done\Phi_{n+2}(p_1+p_2;p_3,p_4,k_1,\dots,k_n)\;
  e^{Y(\Omega;p_1,\dots,p_4)}\;\prod_{i=1}^n\bar\Theta(\Omega;k_i)\\
  &\times\frac14\sum_{\lambda,\sigma_1\dots\sigma_n}\abs{\mathcal M_n\rbr{{}^{p}_{\lambda}\,{}^{k_1}_{\sigma_1}\cdots{}^{k_n}_{\sigma_n}}}^2 .
  \end{aligned}
  \label{eq:master}
\end{equation}
All IR divergences of the processes in \cref{eq:process-n} are contained in the YFS form
factor $Y(\Omega;p_1,\dots,p_4)$, where $\Omega$ is a region surrounding the IR-divergent
soft-photon point $k=0$, and $\bar\Theta(\Omega;k)=1$ for $k\notin\Omega$ and $0$ otherwise.
The amplitudes $\mathcal M_n$ are the UV- and IR-finite $n$-photon spin
amplitudes introduced above, which are built from IR-finite residuals in \cref{subsec:amps}. Finally, $\lambda$
collects the spins of the fermions and $\sigma_i$ the helicities of the photons, the factor $1/4$
averages over the spins of the incoming electron and positron, and $1/n!$ accounts for the $n$
identical photons.

Photons inside the soft region $\Omega$ are therefore not generated: their real emission, summed to
all orders, is combined with the virtual soft corrections at all orders into the
exponential $e^{Y}$, which is then finite by the KLN theorem.\footnote{The same exponentiation underlies the
implementations of refs.~\cite{Jadach:2000ir,Krauss:2022ajk}.}
We define $\Omega$ by the Lorentz-invariant condition
\begin{equation}
  \Omega=\cbr{k:\ \frac{(p_1+p_2)\cdot k}{\sqrt s}<E_{\min}}\,,
  \label{eq:omega}
\end{equation}
which reduces to the per-photon condition $k^0<E_{\min}$ in the centre-of-mass frame of the beams. 
In \cref{sec:low-level-validation}, we verify that physical observables are independent
of sufficiently small $E_{\min}$.

\subsubsection*{The YFS form factor}
%\label{subsec:yfs}

The YFS form factor of \cref{eq:master} can be cast as a sum over the six dipoles formed by
the charged legs
\begin{equation}
  Y(\Omega;p_1,\dots,p_4)=\sum_{i<j}\eta_{ij}\,Y_\Omega(p_i,p_j),\qquad
  \eta_{ij}=-Q_i\theta_i\,Q_j\theta_j\,,
  \label{eq:Ysum}
\end{equation}
so that $\eta_{12}=Q_e^2$,
$\eta_{34}=Q_f^2$, $\eta_{13}=\eta_{24}=+Q_eQ_f$ and $\eta_{14}=\eta_{23}=-Q_eQ_f$.

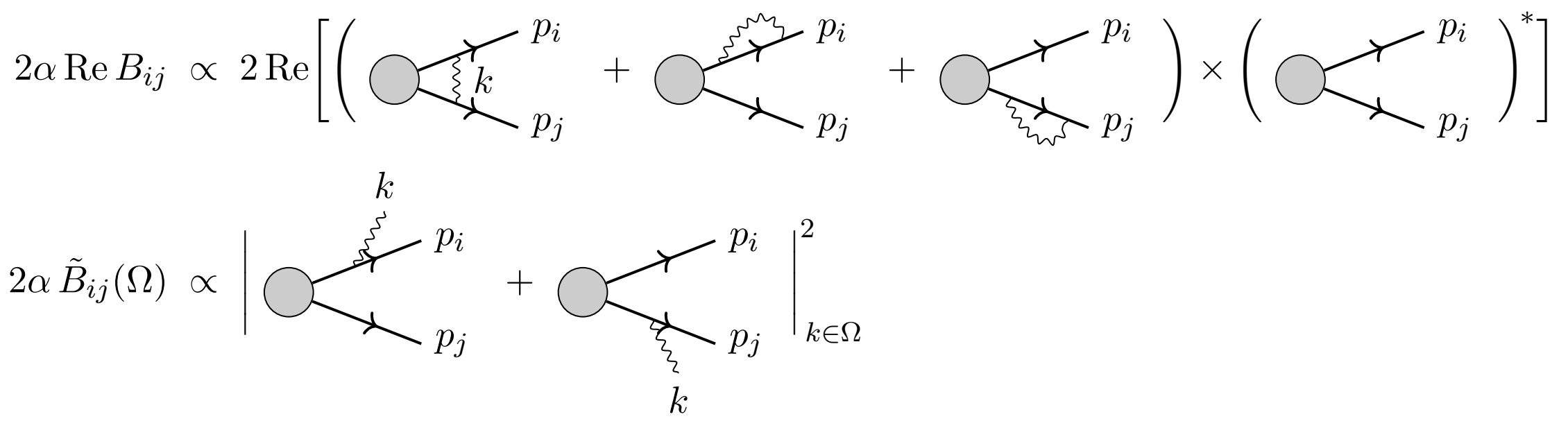
\begin{figure}[t]
  \centering
  \resizebox{\textwidth}{!}{\tikzset{
  photon/.style={decorate,decoration={snake,amplitude=1pt,segment length=3.5pt}},
  leg/.style={thick,postaction={decorate,decoration={markings,mark=at position 0.6 with {\arrow{>}}}}},
  blob/.style={circle,fill=gray!40,draw,minimum size=5mm,inner sep=0pt}
}
\newcommand{\dip}{\node[blob] (b) at (0,0) {};
  \draw[leg] (b) -- (1.8,0.7) node[right] {$p_i$};
  \draw[leg] (b) -- (1.8,-0.7) node[right] {$p_j$};}
\begin{tabular}{@{}r@{\;\;$\propto$\;\;}l@{}}
$2\alpha\,\mathrm{Re}\,B_{ij}$ &
$2\,\mathrm{Re}\Bigg[\Bigg($
\begin{tikzpicture}[baseline=(b.base),scale=0.7]
  \dip \draw[photon] (0.9,0.35) -- node[right=1pt] {$k$} (0.9,-0.35);
\end{tikzpicture}
$\;+\;$
\begin{tikzpicture}[baseline=(b.base),scale=0.7]
  \dip \draw[photon] (0.6,0.23) to[out=70,in=110,looseness=1.6] (1.5,0.58);
\end{tikzpicture}
$\;+\;$
\begin{tikzpicture}[baseline=(b.base),scale=0.7]
  \dip \draw[photon] (0.6,-0.23) to[out=-70,in=-110,looseness=1.6] (1.5,-0.58);
\end{tikzpicture}
$\Bigg)\times\Bigg($
\begin{tikzpicture}[baseline=(b.base),scale=0.7]
  \dip
\end{tikzpicture}
$\Bigg)^{\!*}\Bigg]$
\\[1ex]
$2\alpha\,\tilde B_{ij}(\Omega)$ &
$\Bigg|$
\begin{tikzpicture}[baseline=(b.base),scale=0.75]
  \dip \draw[photon] (0.9,0.35) -- (1.3,1.1) node[above] {$k$};
\end{tikzpicture}
$\;+\;$
\begin{tikzpicture}[baseline=(b.base),scale=0.75]
  \dip \draw[photon] (0.9,-0.35) -- (1.3,-1.1) node[below] {$k$};
\end{tikzpicture}
$\Bigg|^{2}_{\,k\in\Omega}$
\end{tabular}}
  \caption{Soft parts of the squared diagrams constituting the YFS form factor of the dipole $(ij)$: $B_{ij}$ contains the interference of the virtual exchange and of the self-energy-like terms with the Born, and $\tilde B_{ij}$ contains the squared sum of real emissions.}
  \label{fig:yfsdiag}
\end{figure}

For a dipole $(ij)$, its YFS form factor is the sum of a virtual and a real contribution
\begin{equation}
  Y_\Omega(p_i,p_j)=2\alpha\,\mathrm{Re}\,B_{ij}+2\alpha\,\tilde B_{ij}(\Omega)\,,
  \label{eq:Yij-def}
\end{equation}
where the two contributions are given by
\begin{equation}
  \begin{aligned}
  B_{ij}&=-\frac{i}{8\pi^3}\int\frac{\done^4k}{k^2}
  \rbr{\frac{2p_i\theta_i-k}{k^2-2k\cdot p_i\theta_i}+\frac{2p_j\theta_j+k}{k^2+2k\cdot p_j\theta_j}}^{2},\\
  \tilde B_{ij}(\Omega)&=-\frac{1}{8\pi^2}\int_\Omega\frac{\done^3k}{k^0}
  \rbr{\frac{p_i}{k\cdot p_i}-\frac{p_j}{k\cdot p_j}}^{2}.
  \end{aligned}
  \label{eq:BBtilde}
\end{equation}
The diagrams which contribute to each are shown in \cref{fig:yfsdiag}.
The virtual and real contributions are separately IR divergent, with poles of opposite
sign that cancel in their sum. We compute both in dimensional regularisation.\footnote{For the
purposes of this work, all amplitudes are regularised in $D=4-2\epsilon$ dimensions.}
In terms of the dipole invariant $s_{ij}=(p_i+p_j)^2$, their sum reads
\begin{equation}
  Y_\Omega(p_i,p_j)=\frac{\alpha}{4\pi}\,\mathrm{Re}\sbr{c_{ij}\ln\frac{E_{\min}^2}{s_{ij}}+q_{ij}+v_{ij}}\,,
  \label{eq:Yij}
\end{equation}
where $c_{ij}$ is the coefficient of the IR pole, which enters the virtual and real
parts with opposite signs,
\begin{equation}
  \mathrm{Re}\,c_{ij}=4\rbr{\frac{L_{ij}}{\rho_{ij}}-1},\quad
  \rho_{ij}=\frac{\sqrt{P_{ij}^2-m_i^2m_j^2}}{P_{ij}},\quad
  L_{ij}=\ln\frac{P_{ij}(1+\rho_{ij})}{m_im_j},\quad P_{ij}=p_i\cdot p_j\,.
  \label{eq:cij}
\end{equation}
Here $\rho_{ij}$ is the relative velocity of the particles with momenta $p_i$ and $p_j$, while
$v_{ij}$ and $q_{ij}$ are the finite parts of the virtual and real integrals, respectively, given in
\cref{app:yfs}.

We have computed $v_{ij}$ and $q_{ij}$ analytically in the physical region\footnote{
The physical region corresponds to production for the initial--initial and final--final pairs, and to the scattering region for the initial--final pairs.}
and checked them against the
literature~\cite{Passarino:2001wy,Jadach:1995sp,Krauss:2022ajk,Hamilton:2006xz}
(\cref{app:checks}). Since $\Omega$ is defined in \cref{eq:omega} through the
invariant $(p_1+p_2)\cdot k$, $Y_\Omega(p_i,p_j)$ is a Lorentz-invariant function of $p_i$, $p_j$
and $p_1+p_2$: the only reference to a particular frame, the centre-of-mass frame of the beams,
enters through~$\Omega$.

Only the sum of the virtual and real contributions enters the exponent of
\cref{eq:master}. The virtual part alone, however, is also needed to define the one-loop
residuals $\hat\beta^{(1)}_0$ and $\hat\beta^{(2)}_1$ (\cref{subsec:amps,subsec:irsub}), from which
it is removed at the amplitude level. Summing over the dipoles, $B=\sum_{i<j}\eta_{ij}B_{ij}$, we
find
\begin{equation}
  2\alpha B\big|_{\rm fin}=\frac{\alpha}{4\pi}\sum_{i<j}\eta_{ij}\sbr{c_{ij}\ln\frac{\mu^2}{s_{ij}}+v_{ij}}\,,
  \label{eq:BV}
\end{equation}
where $|_{\rm fin}$ denotes the finite part. The full $B$ also contains the IR pole
 of each dipole~\eqref{eq:virtual}, which $e^{-\alpha B}$ in
\cref{eq:Mren} cancels against the IR poles of the bare amplitudes. It is a complex quantity, which contains the Coulomb phases, whereas only
its real part enters $Y$ through \cref{eq:Yij-def}.

\subsubsection*{Amplitudes \& IR finite residuals}
\label{subsec:amps}
The $n$-photon spin amplitude of \cref{eq:master} is a coherent sum over the $2^n$
\emph{partitions}, $\wp\in\{{\rm I},{\rm F}\}^n$, which assign each photon to the initial or
the final state, as follows
\begin{equation}
  \mathcal M_n
  =\sum_{\wp\in\{{\rm I,F}\}^n}\;\prod_{i=1}^{n}\eik^{\wp_i}_{\sigma_i}(k_i)\;
  \sbr{\hat\beta_0(X_\wp)
      +\sum_{j=1}^{n}\frac{\hat\beta^{\wp_j}_{1}(k_j;X_\wp)}{\eik^{\wp_j}_{\sigma_j}(k_j)}
      +\sum_{1\le j<l\le n}\frac{\hat\beta^{\wp_j\wp_l}_{2}(k_j,k_l;X_\wp)}{\eik^{\wp_j}_{\sigma_j}(k_j)\,\eik^{\wp_l}_{\sigma_l}(k_l)}
      +\cdots},
  \label{eq:Mn}
\end{equation}
where the helicity labels of the IR-finite residuals $\hat\beta_m$ are suppressed. Each partition
carries the product of its eikonal factors and the IR-finite residuals; the latter are
evaluated on the reduced kinematics of the partition obtained from the appropriate momentum
mapping (see \cref{subsec:mappings}). In this expression, $X_\wp=p_1+p_2-\sum_{i:\wp_i={\rm I}}k_i$ is the momentum
of the $s$-channel photon of the partition. The residuals are defined recursively from the
amplitudes $\mathcal M^{(n+l)}_n$ with $n$ photons and $l$ loops~\cite{Jadach:2000ir}
  \begin{align}
    \label{eq:betarec}
    \hat\beta^{(l)}_0&=\mathcal M^{(l)}_0\,,
\\
    \hat\beta^{(1+l)}_1(k_1)&=\mathcal M^{(1+l)}_1(k_1)-\hat\beta^{(l)}_0\,\eik(k_1)\,,
\notag\\
    \hat\beta^{(2+l)}_2(k_1,k_2)&=\mathcal M^{(2+l)}_2(k_1,k_2)-\hat\beta^{(1+l)}_1(k_1)\,\eik(k_2)
      -\hat\beta^{(1+l)}_1(k_2)\,\eik(k_1)-\hat\beta^{(l)}_0\,\eik(k_1)\eik(k_2)\,,
      \notag
  \end{align}
where the superscript counts the order in $\alpha$.\footnote{This superscript has been suppressed in
\cref{eq:Mn} in favour of labelling the partition.}
Since $\mathcal M_n$ is renormalised and the virtual IR divergences have been factored out of
it into $B$, the amplitudes $\mathcal M^{(n+l)}_n$ are free of UV and IR
poles, unlike the bare (unrenormalised) amplitudes $\mathcal M_{\text{b},n}$
of \cref{subsec:amplitudes} (see \cref{eq:Mren}). For
$n\ge1$, however, they are still singular when any of their photons becomes soft, as dictated by
\cref{eq:soft-limit}. Each line of \cref{eq:betarec} subtracts these real soft
singularities, given by the lower residuals times eikonal factors, so that the residuals
$\hat\beta_m$ are IR finite and remain finite in the soft limit of any of their photons.
We also note that all amplitudes and eikonal factors are evaluated in one spinor basis, the Kleiss--Stirling spinors~\cite{Kleiss:1985yh} built
from a single reference vector transverse to the beams, $\xi=(1,0,1,0)$, so that the mapped momenta of every
partition share the same phase conventions and the soft limit of \cref{eq:soft-limit} holds
for each helicity.

\begin{table}[!t]
  \centering
  \small
  \setlength{\tabcolsep}{5pt}
  \begin{tabular}{@{}llll@{}}
    \toprule
    process & order & residuals & content \\
    \midrule
    \multirow[t]{4}{*}{$e^+e^-\to f^+f^-$}
      & LO     & $\hat\beta^{(0)}_0$ & Born \\
      & NLO    & $\hat\beta^{(0)}_0,\ \hat\beta^{(1)}_0,\ \hat\beta^{(1)}_1$ & Born, one loop, one photon \\
      & LO+\LLs  & $\hat\beta^{(0)}_0,\ \hat\beta^{(1)}_1,\ \hat\beta^{(2)}_2$ & Born, one and two photons \\
      & NLO+\NLLs & $\hat\beta^{(0)}_0,\ \hat\beta^{(1)}_0,\ \hat\beta^{(1)}_1,\ \hat\beta^{(2)}_2$ & Born, one loop, one and two photons \\
    \midrule
    \multirow[t]{4}{*}{$e^+e^-\to f^+f^-\gamma$}
      & LO     & $\hat\beta^{(1)}_1$ & one photon \\
      & NLO    & $\hat\beta^{(1)}_1,\ \hat\beta^{(2)}_1,\ \hat\beta^{(2)}_2$ & one photon (tree and one loop), two photons \\
      & LO+\LLs  & $\hat\beta^{(1)}_1,\ \hat\beta^{(2)}_2$ & one and two photons \\
      & NLO+\NLLs & $\hat\beta^{(1)}_1,\ \hat\beta^{(2)}_1,\ \hat\beta^{(2)}_2$ & one photon (tree and one loop), two photons \\
    \bottomrule
  \end{tabular}
  \caption{Residuals of \cref{eq:betarec} entering the amplitudes at each order, for
  $f=\mu,\pi$, the superscript counting the order in $\alpha$ and the subscript the number of
  photons, with the two-loop residual $\hat\beta^{(2)}_0$ not included.}
  \label{tab:orders}
\end{table}
\paragraph{Truncation.}
With $\mathcal M_n=\mathcal M^{(0)}_n+\mathcal M^{(1)}_n$ the tree-level and one-loop parts
of the amplitude, the $n$-photon integrand truncated at order $\alpha^{\nmax}$ beyond the Born reads
\begin{equation}
  \begin{aligned}
  \overline{\abs{\mathcal M}^2}_n
  ={}&\frac14\sum_{\lambda,\sigma}\Big[\abs{\mathcal M^{(0)}_n}^2T_{\nmax-n}(Y)
   +2\,\mathrm{Re}\rbr{\mathcal M^{(0)*}_n\mathcal M^{(1)}_n}T_{\nmax-n-1}(Y)\\
   &\qquad\quad+\abs{\mathcal M^{(1)}_n}^2T_{\nmax-n-2}(Y)\Big],
%   \qquad T_m(Y)=\sum_{j=0}^{m}\frac{Y^j}{j!}\,,
  \end{aligned}
  \label{eq:truncation}
\end{equation}
with
\begin{align}
T_m(Y)=\sum_{j=0}^{m}\frac{Y^j}{j!}\,, \qquad\text{with }T_{m<0}=0\,,
\end{align}
and $\nmax$ the highest photon multiplicity generated. Every factor, the
exponential included, is expanded to the same total order $\alpha^{\nmax}$, so the $E_{\min}$
dependence cancels exactly at each $\nmax$ (\cref{subsec:emin}) and $\nmax=1$ ($2$ for
the radiative processes) is the fixed-order NLO integrand term by term. The last term, the
square of the one-loop amplitude, is of order $\alpha^2$ relative to the Born and beyond the NLO
accuracy of this work. It is kept at the resummed orders because the full amplitude
$\mathcal M^{(0)}_n+\mathcal M^{(1)}_n$ is squared, which keeps the integrand positive, and its
truncation factor removes it from the fixed-order integrand.

\Cref{tab:orders} lists, for each order at which the code runs, LO, NLO
and the two resummed orders LO+\LLs and NLO+\NLLs, the residuals of \cref{eq:betarec} that
enter the amplitudes of \cref{eq:Mn}, with up to $\nmax=5$ photons generated by default at the
resummed orders. The two fixed orders are the resummed expansion of \cref{eq:truncation}
truncated at $\nmax=1$ and $\nmax=2$ for the non-radiative and the radiative processes respectively,
with the same amplitudes and the same residuals. For the non-radiative processes the two-photon
residual $\hat\beta^{(2)}_2$ is formally beyond NLO, but it carries the exact two-photon matrix
element and is kept at both resummed orders.
This is so that the tails of the distributions sensitive to
hard real emission, such as the low-mass tail of the charged final state invariant mass, are described by the exact tree rather
than by the product of eikonal factors. The residual is included in the same way in
\KKMC~\cite{Jadach:2000ir} and \Sherpa~\cite{Delorme:2026qbm}.

%\subsection{One-loop amplitudes}
\subsection{Scattering amplitudes to one loop}
\label{subsec:amplitudes}

For the tree and one-loop amplitudes of the energy-scan and radiative-return processes, we do
not rely on automated tools. Instead, we construct our own amplitudes, optimised for our in-house
implementation of CEEX (see \cref{subsec:implementation}). We compute them with modern
techniques for the analytic calculation of scattering amplitudes and the fast numerical evaluation
of Feynman integrals, described below.

Constructing our own amplitudes has two advantages. First, it provides an independent
calculation of the virtual corrections, which is a check of the fixed-order predictions in its own
right: every tree and one-loop amplitude has been verified against \GoSam~\cite{GoSam:2014iqq},
with \Ninja~\cite{Peraro:2014cba} or \Golem~\cite{Binoth:2008uq} reduction, and the trees also
against the Feynman diagrams evaluated in high precision. It also allows us to optimise the
analytic expressions with \FORM~\cite{Vermaseren:2000nd,Kuipers:2012rf,Ruijl:2017dtg} and compile them into Fortran libraries, which makes
their evaluation fast enough for the resummation within a standalone implementation. Second, it
gives us control over the analytic structure of the amplitudes, in particular over the phases of
the spinors. 
The residuals of \cref{eq:betarec} are differences between amplitudes with
different numbers of photons, and the eikonal factors of \cref{eq:eikonal} multiply
amplitudes with one photon fewer, all evaluated on the mapped momenta of
\cref{subsec:mappings}. These combinations have the correct soft limit only if every
amplitude is built on the same spinors, with the same phase conventions, which no external
library guarantees. 

With our own amplitudes, the soft-limit relation of \cref{eq:soft-limit}
can be checked directly at the amplitude level, helicity by helicity, and the residuals are free
of spurious phases. In the following, we describe how the tree and one-loop amplitudes of each
process are computed, and then their IR subtraction and renormalisation.

\subsubsection*{Tensor decomposition and diagram-by-diagram evaluation}

The one-loop amplitudes of $e^+e^-\to\mu^+\mu^-$, $e^+e^-\to\pi^+\pi^-$ and
$e^+e^-\to\pi^+\pi^-\gamma$ are computed analytically with the four-dimensional tensor
decomposition of refs.~\cite{Peraro:2019cjj,Peraro:2020sfm,Chen:2019wyb,Dave:2024ewl}, following
ref.~\cite{Dave:2026pvq}. Each amplitude is written as a set of form factors multiplying
independent tensor structures, listed in \cref{app:tensors} for the scan processes, with the external states kept in four dimensions while the loop
momenta live in $D=4-2\epsilon$ dimensions, and all one-loop integrals are evaluated numerically
with \Collier~\cite{Denner:2016kdg}. 
The one-loop amplitude of $e^+e^-\to\mu^+\mu^-$ is obtained from four Feynman diagrams
and can be expressed in terms of eight form factors, whose Feynman integrals are reduced to 17 master
integrals. For $e^+e^-\to\pi^+\pi^-$, with the pions treated as point-like particles in scalar
QED, seven diagrams give two form factors and 14 master integrals. For
$e^+e^-\to\pi^+\pi^-\gamma$, we use the polarised amplitudes of ref.~\cite{Dave:2026pvq}, eight
form factors from 61 diagrams in six gauge-invariant groups, from which we remove the pion
bubble on the $s$-channel photon, the scalar-QED contribution to the hadronic VP,
since the latter is already contained in the dressed form factor and no fermion loop is inserted
on a photon propagator in any of our amplitudes.

The one-loop amplitude of $e^+e^-\to\mu^+\mu^-\gamma$ is instead evaluated numerically,
diagram by diagram, without any prior algebraic manipulation: the numerator of each diagram is a
polynomial in the loop momentum built from the spinors of the tree amplitudes, and it is
contracted with the tensor integrals of \Collier, up to rank three for the pentagons. The 32 diagrams, without self-energies
on the external legs and without VP, form the complete set of NLO QED diagrams (see
the discussion in ref.~\cite{Campanario:2013uea}).
%The rational terms of the 't~Hooft--Veltman scheme are local
%constants for the vertex and self-energy insertions and vanish for the boxes.}

Finally, the tree amplitudes with one and two photons, which enter the residuals
$\hat\beta_1$ and $\hat\beta_2$, are evaluated for every helicity and every partition,
separately for initial- and final-state emission.

In both pion processes, the pion form factor $F_\pi$ multiplies the tree-level and the one-loop
amplitudes alike, at the virtuality $s$ of the $s$-channel photon in the scan mode and, in the
radiative return, at the scale $X_\wp^2$ of each partition $\wp$ of \cref{eq:Mn}. This is the
factorised approach usually referred to as F$\times$sQED, and is implemented by all the generators we compare with in
\cref{sec:results}.

\subsubsection*{IR subtraction and renormalisation}
\label{subsec:irsub}

The one-loop residuals of \cref{eq:betarec} are built from the bare one-loop amplitudes
$\mathcal M_{\text{b},n}$ described above, computed in dimensional regularisation to treat
UV and IR divergences. The amplitudes $\mathcal M_n$ of
\cref{eq:master} are obtained from them by adding the on-shell counterterms $\delta\mathcal M_n$
and by removing the virtual form factor $\alpha B$ of \cref{eq:BV} at the amplitude level
\begin{equation}
  \mathcal M_n=\rbr{\mathcal M_{\text{b},n}+\delta\mathcal M_n}e^{-\alpha B}\,.
  \label{eq:Mren}
\end{equation}
The bare one-loop amplitudes contain both UV and IR poles in $\epsilon$. The
counterterms $\delta\mathcal M_n$ cancel the UV poles, and the factor $e^{-\alpha B}$ the
remaining IR ones, including those of the wave-function constants below, so that $\mathcal M_n$
is finite. At tree level there are no poles,
$\mathcal M^{(n)}_n=\mathcal M^{(n)}_{\text{b},n}$, so that, expanding \cref{eq:Mren} to first
order in $\alpha$, the one-loop residuals read
\begin{equation}
  \begin{aligned}
  \hat\beta^{(1)}_0&=\mathcal M^{(1)}_{\text{b},0}{}+\delta\mathcal M^{(1)}_0-\alpha B\,\hat\beta^{(0)}_0\,,\\
  \hat\beta^{(2)}_1(k)&=\sbr{\mathcal M^{(2)}_{\text{b},1}(k){}+\delta\mathcal M^{(2)}_1(k)-\alpha B\,\mathcal M^{(1)}_1(k)}-\hat\beta^{(1)}_0\,\eik(k)\,,
  \end{aligned}
  \label{eq:beta-oneloop}
\end{equation}
where $\mathcal M^{(0)}_0$ and $\mathcal M^{(1)}_1(k)$ are the tree-level amplitudes with zero
and one photon, and the counterterms are, in the conventions of ref.~\cite{Dave:2026pvq},
\begin{equation}
  \begin{aligned}
  \delta\mathcal M^{(1)}_0&=\frac{\alpha}{4\pi}\sbr{\delta Z_F(m_e)+\delta Z_f}\mathcal M^{(0)}_0\,,\\
  \delta\mathcal M^{(2)}_1(k)&=\frac{\alpha}{4\pi}\Big\{\sbr{\delta Z_F(m_e)+\delta Z_f}\mathcal M^{(1)}_1(k)
  +\delta m_e\,\mathcal M^{(1)}_{1,\delta m_e}(k)+\delta m_f\,\mathcal M^{(1)}_{1,\delta m_f}(k)\Big\}\,.
  \end{aligned}
  \label{eq:ren}
\end{equation}
Three kinds of counterterms are needed. The
complete on-shell wave-function constant of a fermion of mass $m$, IR
part included, $\delta Z_F(m)=-\rbr{3/\epsilon+3\ln(\mu^2/m^2)+4}$, enters for the electron line in all four processes and for the muon line in the
muon processes, $\delta Z_f=\delta Z_F(m_\mu)$, while for the pion it vanishes in our scheme,
$\delta Z_\pi=0$.\footnote{We keep the IR part of the wave-function constants, whose pole
cancels against the self-energy-like terms of $B$ (see \cref{app:yfs}); for the pion, the UV
and IR parts are equal and opposite. In the Catani--Dittmaier--Tr\'ocs\'anyi
subtraction~\cite{Catani:2000ef}, used for instance in ref.~\cite{Dave:2026pvq}, only the UV part is
kept, so that $\delta Z_\pi=2\rbr{1/\epsilon+\ln(\mu^2/m_\pi^2)+2}$. The two schemes agree once each
is combined with its own IR subtraction, and ref.~\cite{Delorme:2026qbm} follows the same approach
as ours.} The purely UV on-shell mass counterterms
$\delta m_e$ and $\delta m_f$, with $f=\mu$ or $\pi$, enter only the radiative processes,
through the tree-level one-photon amplitudes $\mathcal M^{(1)}_{1,\delta m}(k)$ with the
counterterm inserted on the internal propagator of the emitting line. The photon self-energy is
not included, so that $\alpha=\alpha(0)$ is not renormalised and no photon wave-function
constant appears. Once the counterterms are added, the remaining poles of the amplitude are
IR and cancel analytically against the pole of $B$, a cancellation we check numerically at
every phase-space point.

%==================================================================

\subsection{Momentum mappings}
\label{subsec:mappings}

In order to effectively match fixed-order matrix elements with the soft-photon
resummation, a map from multi-photon emission configurations to each
fixed-multiplicity configuration must be defined.
Each event has a full multi-photon phase-space $\Phi_{4+n_\gamma}$, which must be
mapped to lower-dimensional phase-spaces, $\Phi_{4+N}$, where $N$ is the
number of additional photons in the matrix element being evaluated.
These mappings are not uniquely defined~\cite{Jadach:1988gb}, so here we choose
a mapping procedure similar to that described in ref.~\cite{Price:2025fiu}, but modified to
the case of CEEX.

We start by noting that in the CEEX procedure, for each partition each photon
has a definite assignment $\wp_i={\rm I}$ or ${\rm F}$. Thus, we decompose
$N = N_{\rm I}+N_{\rm F}$. Clearly, a physically motivated
mapping procedure will produce different mapped momenta for each partition.
The momentum mapping never changes the physical momenta of the event, only
the momenta which enter the evaluation of the regularised amplitudes.

For the initial-state mapping, we begin with a list of photons which are assigned to the
initial state in this partition, $\{ k_{\rm I} \}$, and a sum of matched
initial-state photon momenta $K_{\rm I} = \sum_{l=1}^{N_{\rm I}} k_l$.
The unmapped beam momenta before any emission are denoted $\ptilde{1}, \ptilde{2}$.
First we define a frame $Q$
\begin{equation}
  Q = \ptilde{1} + \ptilde{2} + \sum_{\rm I} k_{\rm I} - K_{\rm I}\,,
\end{equation}
which is related to the mapped initial-state momenta $p_1, p_2$ as $Q = p_1+p_2-K_{\rm I}$.
We boost the spacelike vector $\ptilde{1}-\ptilde{2}$ into the frame $Q$ to provide a
direction reference for the lab-frame $z$ axis.

Next, we construct the mapped initial state back-to-back in the frame $Q$ as required
by momentum conservation
\begin{equation}
  p_1 = (\sqrt{m_1^2+p_z^2}, 0, 0, p_z)\,, \qquad p_2 = (\sqrt{m_2^2+p_z^2}, 0, 0, -p_z)\,,
\end{equation}
where
\begin{equation}
  p_z = \frac{1}{2} \sqrt{\frac{\lambda(Q^2,m_1^2,m_2^2)}{Q^2}}\,,
\end{equation}
with $\lambda(a,b,c)$ the K\"all\'en function.

Finally, we undo the rotation and boost the new $p_1, p_2$ back to the original
lab frame. The configuration now conserves momentum when all unmatched photons are
ignored, with all particles still on their mass shells. The initial-state momenta
are no longer necessarily aligned with the $z$ axis.\footnote{Since the amplitudes are summed before squaring and their phases depend on the
frame in which the helicities are defined, it is not possible to use a different Lorentz frame
for each residual.}
Note that all Born-level final-state momenta and all final-state photons, as well as the
matched initial-state photons, are untouched by this mapping procedure.

The final-state mapping follows a similar formulation. The procedure is simplified
by the fact that we only consider processes with two charged particles in the final
state, so that there is only one final-state dipole throughout the calculation.
We again define a sum of matched photon momenta, $K_{\rm F} = \sum_{m=1}^{N_{\rm F}} k_m$,
and a list of all photons which belong to the final-state radiation in this partition,
$\{k_{\rm F}\}$.
The charged legs' momenta before the mapping are denoted $\ptilde{3}, \ptilde{4}$ in analogy
to the initial-state mapping. We define a frame
\begin{equation}
  Q' = \ptilde{3}+\ptilde{4} + \sum_{\rm F} k_{\rm F} -K_{\rm F}\,,
\end{equation}
and boost $\ptilde{3}-\ptilde{4}$ to this frame to define the dipole axis. In the frame $Q'$,
we then define the new dipole momenta back-to-back as
\begin{equation}
  p_3 = (\sqrt{m_3^2+q_z^2}, 0, 0, q_z)\,, \qquad p_4 = (\sqrt{m_4^2+q_z^2}, 0, 0, -q_z)\,,
\end{equation}
where
\begin{equation}
  q_z = \frac{1}{2} \sqrt{\frac{\lambda(Q'^2,m_3^2,m_4^2)}{Q'^2}}\,,
\end{equation}
such that $p_3+p_4=Q'$ as required by momentum conservation.
Finally, we boost and rotate everything back to the lab frame.

\begin{figure}[!htb]
  \centering
  \includegraphics[width=\textwidth]{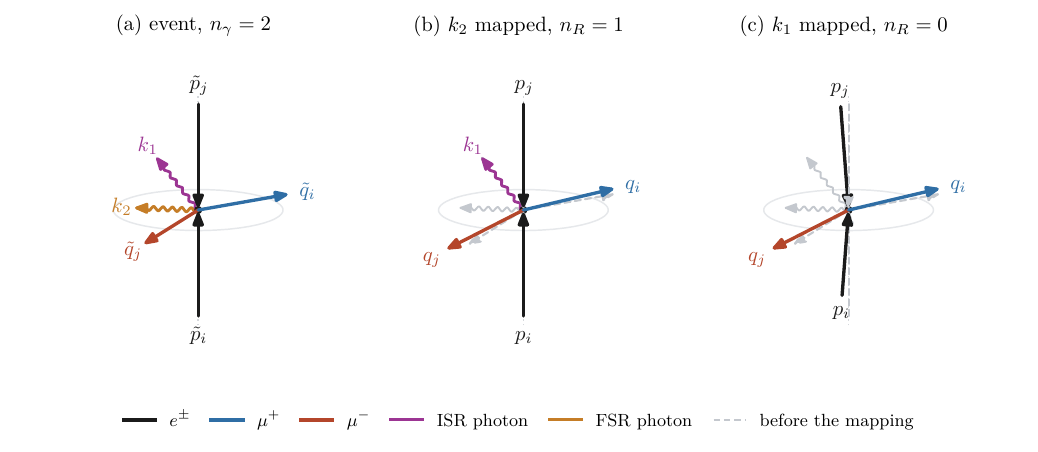}
  \caption{Momentum mappings of one partition of an $e^+e^-\to\mu^+\mu^-\gamma\gamma$ event in
  which $k_1$ is assigned to the initial state and $k_2$ to the final state: (a)~the physical momentum configuration of the event,
  (b)~the configuration after the final-state mapping with no matched photons, which absorbs $k_2$ into the muon pair and keeps $k_1$, and
  (c)~after the initial-state mapping with no matched photons, which absorbs $k_1$ into the beams. In each case the dashed grey lines show the event before the mapping.}
  \label{fig:mapping}
\end{figure}

For each $\hat\beta$ evaluation, we first apply the final-state mapping, then
the initial-state mapping, to obtain the $(2+N)$-body configuration, as illustrated in
\cref{fig:mapping} for one partition of a two-photon event.
These mappings are soft- and collinear-safe and do not affect the logarithmic accuracy of the
resummation.

%====================================================================

\subsubsection*{Generation of multi-photon phase-space} \label{sec:ps}

The main strength of Monte Carlo event generators is the ability to randomly sample from probability distributions to produce fully differential simulations. We employ the well-known technique of multi-channel sampling to efficiently draw from the multi-photon phase-space distribution.
We also provide an implementation of \Vegas~\cite{Lepage:1977sw,Lepage:1980dq} as an alternative.

We first recall how Monte Carlo sampling works in one dimension. Given a probability distribution $f(x)$ from which we would like to randomly sample a value $x$, and a uniform random number $R \in (0,1)$, we may generate $x$ using
\begin{equation} \label{eq:MCsampling}
  x = F^{-1}(F(\xmin) + R(F(\xmax)-F(\xmin))),
\end{equation}
where $[\xmin,\xmax]$ is the allowed range of $x$, $F(x)$ is the primitive function of $f$, and $F^{-1}$ is its inverse. However, in most applications, the probability distribution $f(x)$ (corresponding to a differential cross-section or decay rate) cannot be analytically integrated. Instead, we can achieve efficient sampling by identifying a simple function $g(x)$ such that $f(x) \leq g(x)$ in the integration region, and $G(x)$ and its inverse are known. Then, after generating $x$ according to the distribution $G(x)$ (as in \cref{eq:MCsampling}), the point is given a weight $f(x)/g(x)$. This ensures that all differential and integrated quantities carry the correct weight.\footnote{To generate physical events, an \emph{unweighting} step must be added which carries out a hit-and-miss sampling of events according to their total weight, which includes both the phase-space weight and the amplitude.}

The multi-channel sampling technique is to extend this method by choosing $g(x)$ according to a sum of physically motivated \emph{channels} $g_i(x)$. Below we outline our choices of the most relevant channels for generating multiple photons. All channels are listed in \cref{app:channels}.

First, each event has a number of photons $n_\gamma$, which is generated according
to a Poisson distribution with mean
\begin{equation}
  \langle n_\gamma \rangle = \frac{2\alpha}{\pi} \left(\ln\left(\frac{s}{m_e^2}\right) - 1 + \ln\left(\frac{s}{m_f^2}\right) - 1\right) \ln\left(\frac{E_{\max}}{E_{\min}}\right),
  \label{eq:nphoton}
\end{equation}
with $E_{\rm max}$ the maximum possible photon energy. Once the multiplicity of the event is fixed, the photons' energies and angles must be sampled.\footnote{The energy distribution and angle distributions are factorised, and the azimuthal angle is always uniformly sampled.}

The photon energies are sampled according to
\begin{equation*}
g_{\rm soft}(E) = \frac{1}{E \,\ln(E_{\max}/E_{\min})}, \quad \text{giving} \quad
E = E_{\min}\left(\frac{E_{\max}}{E_{\min}}\right)^{R},
\end{equation*}
where $E_{\min}$ and $E_{\max}$ depend on the experimental scenario and selection cuts.
In particular, for radiative return, one photon is generated with higher energy. In the case of radiative pion production, an additional photon-energy channel is added due to the $\rho$ resonance (see \cref{app:channels}).

The angular distribution of photons is given by
\begin{equation}
g_{\rm angle}(\hat k) = \frac{c_0}{4\pi} + c_e \frac{\beta_e}{4\pi \,\artanh\beta_e}\,\frac{1}{1 - \beta_e^2\cos^2\theta} + c_f \frac{\beta_f}{4\pi \,\artanh\beta_f}\,\frac{1}{1 - \beta_f^2\cos^2\theta_{kf}},
\end{equation}
where the $c_0, c_e, c_f$ are coefficients denoting the relative balance of flat, initial-state-collinear and final-state-collinear angular enhancements. Their exact assignment differs depending on the scenario and cuts. In radiative-return processes, the photon generated with higher energy has larger $c_0$, such that its distribution is wide-angle.

\subsection{Structure of the code}
\label{subsec:implementation}

% Shared styles of the code-map figures.
\tikzset{
  cmap/.style={font=\small},
  box/.style={draw,rounded corners=2pt,thick,align=center,font=\small,inner sep=4pt,fill=white},
  layer/.style={draw,thick,rounded corners=3pt,inner sep=6pt,fill=gray!8},
  srv/.style={box,fill=blue!8},
  amp/.style={box,fill=orange!12},
  lib/.style={box,fill=green!12},
  ext/.style={box,fill=gray!20,dashed},
  flow/.style={-{Stealth[length=5pt]},thick},
  dep/.style={-{Stealth[length=4pt]},thin,gray!70!black},
  lbl/.style={font=\footnotesize\itshape,gray!60!black}}

\begin{figure}[t]
  \centering
  \resizebox{0.95\textwidth}{!}{% Path of one event through the code. Boxes white; the three stages are shaded backgrounds.
\definecolor{stgen}{HTML}{1F3A7A}
\definecolor{stamp}{HTML}{5A5A5A}
\definecolor{sthis}{HTML}{2E6B4F}
\begin{tikzpicture}[cmap,x=1cm,y=1cm,
  blk/.style={draw=black!70,rounded corners=2pt,thick,align=center,font=\small,inner sep=5pt,fill=white},
  stagelbl/.style={font=\footnotesize\scshape,rotate=90,anchor=north,inner sep=2pt,align=center}]
  % ---- event generation
  \node[blk,minimum width=12.4cm] (smp)  at (0,1.1)  {\textbf{sampling} of the pair and photon momenta};
  \node[blk,minimum width=12.4cm] (ev)   at (0,0.0)  {\textbf{event} inside the cuts of the scenario};
  % ---- amplitude
  \node[blk,minimum width=12.4cm] (part) at (0,-1.5) {\textbf{partitions} of the photons between initial and final state};
  \node[blk,minimum width=3.9cm,minimum height=1.0cm] (map) at (-4.25,-2.8) {\textbf{mappings}};
  \node[blk,minimum width=3.9cm,minimum height=1.0cm] (eik) at (0,-2.8) {\textbf{eikonal factors}};
  \node[blk,minimum width=3.9cm,minimum height=1.0cm] (res) at (4.25,-2.8) {\textbf{residuals}};
  \node[blk,minimum width=12.4cm] (sum) at (0,-4.1) {\textbf{construction of the amplitude}, coherent sum over the partitions};
  \node[blk,minimum width=12.4cm] (sq)  at (0,-5.2) {\textbf{squared amplitude} with the truncated YFS form factor};
  % ---- analysis
  \node[blk,minimum width=3.9cm,minimum height=1.0cm] (obs) at (-4.25,-6.7) {\textbf{observables}};
  \node[blk,minimum width=3.9cm,minimum height=1.0cm] (his) at (0,-6.7) {\textbf{histograms}};
  \node[blk,minimum width=3.9cm,minimum height=1.0cm] (out) at (4.25,-6.7) {\textbf{cross-section}};
  % ---- stage backgrounds (drawn behind, same width, same left edge)
  \begin{scope}[on background layer]
    \fill[stgen!10,rounded corners=4pt] (-7.3,1.6) rectangle (6.6,-0.5);
    \fill[stamp!10,rounded corners=4pt] (-7.3,-0.95) rectangle (6.6,-5.7);
    \fill[sthis!10,rounded corners=4pt] (-7.3,-6.0) rectangle (6.6,-7.4);
  \end{scope}
  \node[stagelbl,text=stgen] at (-7.2,0.55) {event\\generation};
  \node[stagelbl,text=stamp] at (-7.2,-3.3) {amplitude};
  \node[stagelbl,text=sthis] at (-7.2,-6.7) {analysis};
\end{tikzpicture}}
  \caption{Path of one event through the code, from the phase-space sampling to the coherent sum over the partitions, the squared amplitude with the YFS form factor, and the analysis.}
  \label{fig:codemap}
\end{figure}
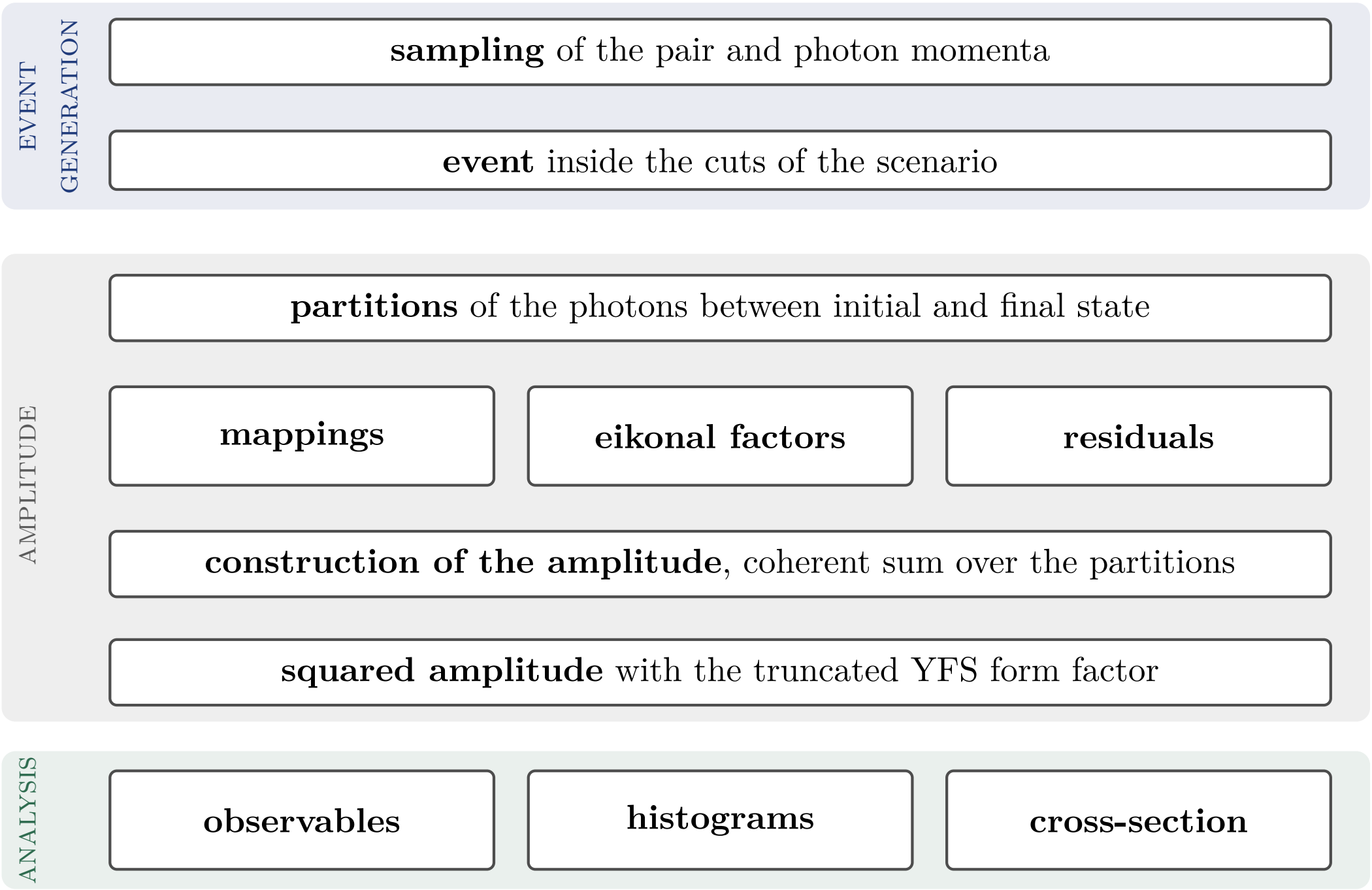

We have implemented the methods described above in an in-house Monte Carlo integrator written
in {\sc Fortran}, which produces the integrated cross-section together with histograms of the
requested differential distributions, and which will be merged into the \Phokhara code.
The program is organised in three stages, sketched in \cref{fig:codemap}.
The generation stage draws the momenta of the final-state charged pair and of the photons
from the multi-channel sampler of \cref{sec:ps}, or from \Vegas~\cite{Lepage:1977sw,Lepage:1980dq},
applying the relevant cuts. The amplitude stage assembles, for each partition, the mapped
momenta of \cref{subsec:mappings}, the eikonal factors and the residuals, sums them
coherently into the $n$-photon amplitude of \cref{eq:Mn} and squares it with the truncated
form factor of \cref{eq:truncation}. The code can also be run in fixed-order mode, which uses the same residuals. Finally, the
analysis stage fills the requested histograms and accumulates the cross-section.
% The ingredients shared by every process are the eikonal factors, the YFS form factor of
% \cref{app:yfs}, the momentum mappings, the spinors and polarisation vectors, the
% counterterms of \cref{app:renorm} and the pion form factor, the latter in the
% parametrisation shared by the generators of ref.~\cite{Aliberti:2024fpq}.

\section{Consistency checks}
\label{sec:low-level-validation}

Before comparing against other codes and methods, we first validate the ingredients of
\cref{sec:methods}. We present checks of the cancellation of the poles of the one-loop amplitudes against the counterterms and the YFS virtual factor, we show the independence of the predictions from the soft-photon cut $E_{\min}$, and we validate our fixed-order NLO predictions against independent calculations. Finally, we illustrate the impact of initial--final interference in the CEEX formalism.

In this section and in \cref{sec:results}, we use five scenarios inspired by low-energy experiments at which the $e^+e^- \to \mu^+\mu^-$ and $e^+e^-\to \text{hadrons}$ cross-sections have been measured. These fall into two categories: energy-scan experiments, in which the collider centre-of-mass energy is varied to obtain bins in $\sqrt{s}$, and radiative-return experiments, which fix the collider energy and instead rely on QED initial-state radiation to lower the effective centre-of-mass energy~\cite{Binner:1999bt,Benayoun:1999hm}. In both cases, the experimental analyses rely on Monte Carlo generators to correct for radiative effects, and in some cases also for the normalisation of the cross-section. We use a scan-inspired scenario, CMD~\cite{CMD-3:2023alj}, choosing one collider energy $\sqrt{s}=0.7$ GeV, and four radiative-return scenarios inspired by KLOE~\cite{KLOE-2:2017fda}, BES\,III~\cite{BESIII:2015equ}, and B factories such as BaBar~\cite{BaBar:2012bdw} and Belle. The KLOE small-angle and large-angle selection cuts, as used in the KLOE08~\cite{KLOE:2008fmq} and KLOE12~\cite{KLOE:2012anl} analyses and in the KLOE10~\cite{KLOE:2010qei} analysis, respectively, are treated in two different scenarios.
The selection cuts for each scenario are detailed in \cref{sec:results}.

\subsection{Cancellation of poles}
\label{subsec:pole}

\begin{figure}[t]
  \centering
  \includegraphics[width=0.62\textwidth]{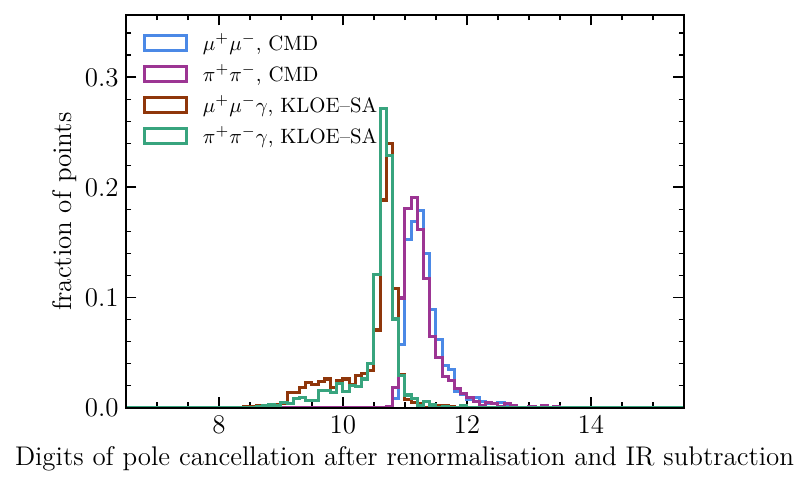}
  \caption{Number of digits to which the $1/\epsilon$ pole of the virtual correction cancels after
  renormalisation and IR subtraction, for $\mu^+\mu^-$ and $\pi^+\pi^-$ in
  the CMD scenario and $\mu^+\mu^-\gamma$ and $\pi^+\pi^-\gamma$ in the KLOE small-angle scenario,
  with point-like pions. The number of digits is calculated as
  $-\log_{10}\sbr{\abs{V+M+F-2\alpha B}/\max\rbr{\abs V,\abs M,\abs F,\abs{2\alpha B}}}$ (definitions given in main text), evaluated at 2000 accepted phase-space points per process.}
  \label{fig:pole}
\end{figure}

The UV and IR poles of the bare one-loop amplitude must cancel against the counterterms and the virtual form
factor $\alpha B$ point by point, \cref{eq:Mren}. This is checked for every event; \cref{fig:pole} shows
the number of digits to which the cancellation of the $1/\epsilon$ coefficient occurs, for 2000
phase-space points per process inside the acceptance of the CMD and KLOE small-angle scenarios, the same check having been performed in the other scenarios.
\Cref{tab:pole} lists the individual contributions at one phase-space point per process: the
bare one-loop amplitude $V$, the sum $M+F$ of the mass and field counterterms, and the form factor term $-2\alpha B$, at the phase-space points listed in \cref{app:pole-points}.
\begin{table}[!htb]
  \centering
  \footnotesize
  \setlength{\tabcolsep}{3pt}
  \begin{tabular}{llrrrr}
    \toprule
    process & scenario & one loop $V$ & counterterms $M+F$ & $-2\alpha B$ & sum \\
    \midrule
    $\mu^+\mu^-$ & CMD & $\phantom{-}72.190087992667$ & $-12.000000000000$ & $-60.190087992820$ & $-1.5\times 10^{-10}$ \\
    $\pi^+\pi^-$ & CMD & $\phantom{-}64.025195122210$ & $-6.000000000000$ & $-58.025195122583$ & $-3.7\times 10^{-10}$ \\
    $\mu^+\mu^-\gamma$ & KLOE small-angle & $\phantom{-}76.440356253431$ & $-12.366977534501$ & $-64.073378717359$ & $1.6\times 10^{-9}$ \\
    $\pi^+\pi^-\gamma$ & KLOE small-angle & $\phantom{-}68.053162452607$ & $-6.178346766452$ & $-61.874815684922$ & $1.2\times 10^{-9}$ \\
    \bottomrule
  \end{tabular}
  \caption{The $1/\epsilon$ coefficient of the virtual correction, interfered with the tree and
  summed over helicities, in units of $\alpha/(4\pi)$ times the squared tree-level amplitude, at
  the first accepted point of the fixed-seed sample of \cref{fig:pole} for each process: the
  bare one loop $V$, the mass and field counterterms $M+F$, the YFS virtual factor $-2\alpha B$,
  and their sum, at the points listed in \cref{app:pole-points}.}
  \label{tab:pole}
\end{table}
The IR part of the
pole is the one predicted by the universal soft anomalous dimension of
ref.~\cite{Catani:2000ef}, which the form factor reproduces by construction.

\subsection{Independence of the soft cut}
\label{subsec:emin}

\begin{table}[t]
  \centering
  \footnotesize
  \begin{tabular}{lcr}
    \toprule
    $E_{\min}/\sqrt s$ & $\sigma$ [nb] & $\delta\ [10^{-4}]$ \\
    \midrule
    $10^{-2}$ & $573.006(17)$ & $+99.57$ \\
    $10^{-3}$ & $567.924(22)$ & $+10.00$ \\
    $10^{-4}$ & $567.383(30)$ & $+0.46$ \\
    $10^{-5}$ & $567.320(39)$ & $-0.64$ \\
    \midrule
    \BabaYaga & $567.3568(16)$ & \\
    \bottomrule
  \end{tabular}
  \caption{Total cross-section of $e^+e^-\to\pi^+\pi^-$ at NLO in the CMD scenario,
  $\sqrt s=0.7$~GeV, as a function of the soft cut $E_{\min}$ in units of $\sqrt s$, with
  $\delta$ its relative difference to \BabaYaga at NLO without VP.}
  \label{tab:emin-cmd-nlo-pi}
\end{table}

The region $\Omega$ of \cref{eq:omega} splits the photon phase-space according to a parameter $E_{\min}$, the soft cut. Photons
below the cut are integrated analytically into the YFS form factor whilst photons above it are explicitly
generated, and the cross-section must not depend on $E_{\min}$ once it is sufficiently small.
\Cref{tab:emin-cmd-nlo-pi} shows this for $e^+e^-\to\pi^+\pi^-$ at fixed-order NLO in the CMD
scenario, where the same cut acts as the slicing parameter of the real emission: as $E_{\min}$
decreases, the total cross-section converges to that of \BabaYaga, with which it agrees within
$10^{-4}$ for $E_{\min}/\sqrt s\le10^{-4}$. The same test has been performed for the other scenarios, with the
same outcome.

\begin{figure}[t]
  \centering
  \includegraphics[width=0.49\textwidth]{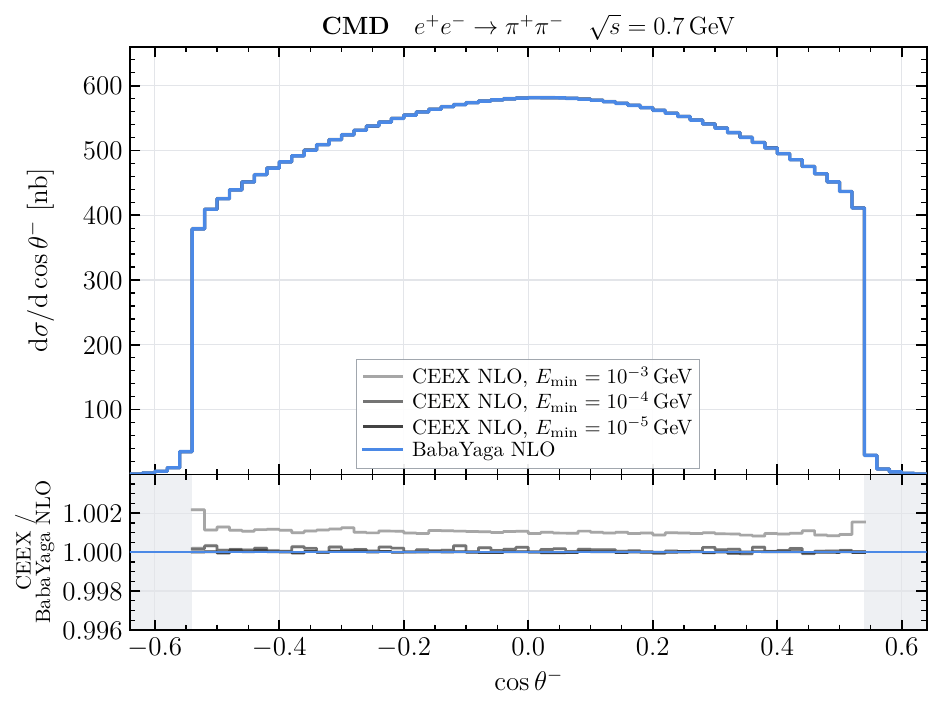}
  \includegraphics[width=0.49\textwidth]{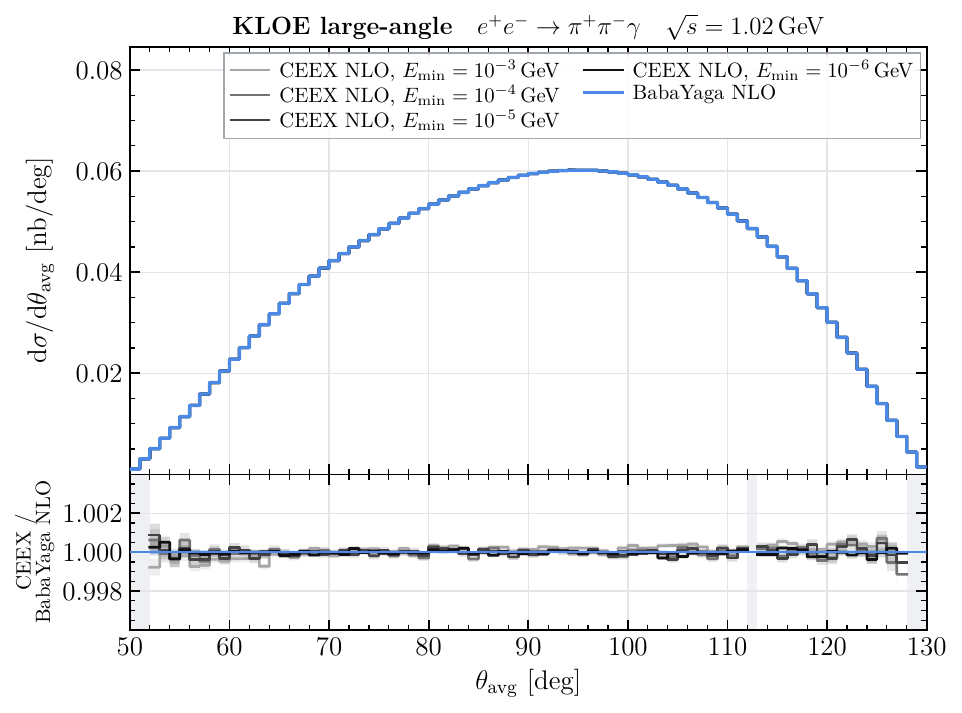}
  \caption{Polar-angle distribution of the $\pi^-$ in $e^+e^-\to\pi^+\pi^-$ with the
  CMD selection (left) and average polar angle of the final-state charged pair in $e^+e^-\to\pi^+\pi^-\gamma$ with the
  KLOE large-angle selection (right) at fixed-order NLO, for $E_{\min}$ from $10^{-3}$~GeV (light
  grey) to $10^{-5}$ and $10^{-6}$~GeV (black), with the ratio to \BabaYaga at NLO in the lower
  panel.}
  \label{fig:emin-cmd-nlo-pi}
\end{figure}

At the differential level, \cref{fig:emin-cmd-nlo-pi} shows the polar-angle distribution of
the $\pi^-$ at CMD and the average polar angle of the final-state charged pair in the KLOE large-angle selection,
for $E_{\min}$ between $10^{-3}$ and $10^{-6}$~GeV. At CMD the predictions converge for
$E_{\min}=10^{-4}$~GeV and below, where they agree with \BabaYaga to better than a per mille over
the whole acceptance, while $E_{\min}=10^{-3}$~GeV is still 1 per mille too high. In the
radiative return the dependence is weaker, the four curves agreeing with \BabaYaga within a per
mille.

\subsection{Fixed-order NLO}
\label{subsec:nlo}
\begin{figure}[!tb]
  \centering
  \raisebox{-\height}{\includegraphics[width=0.49\textwidth]{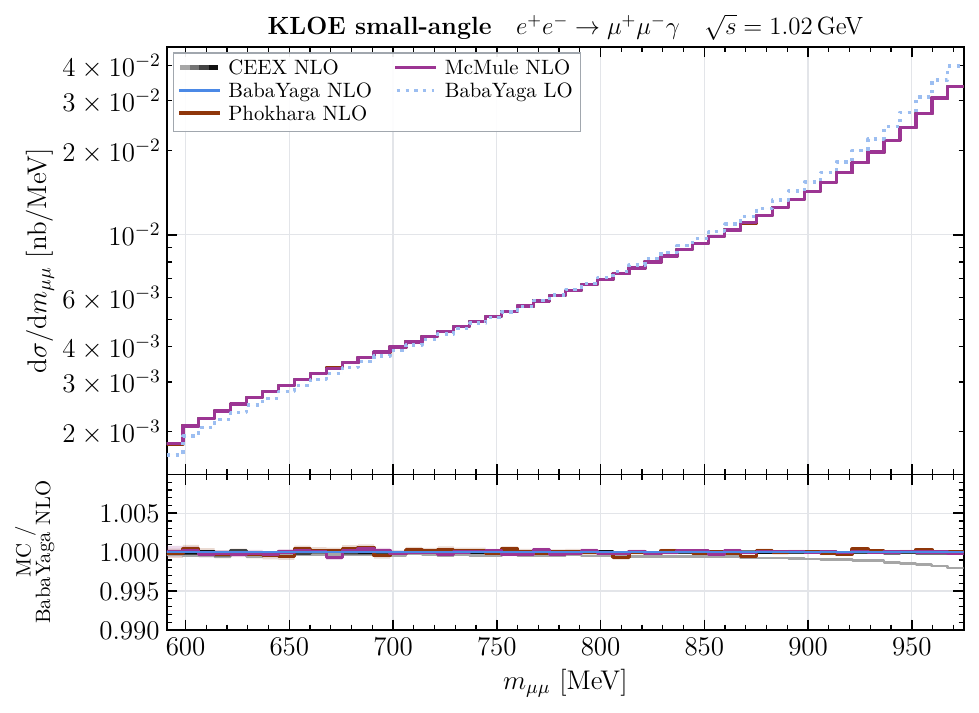}}
  \raisebox{-\height}{\includegraphics[width=0.49\textwidth]{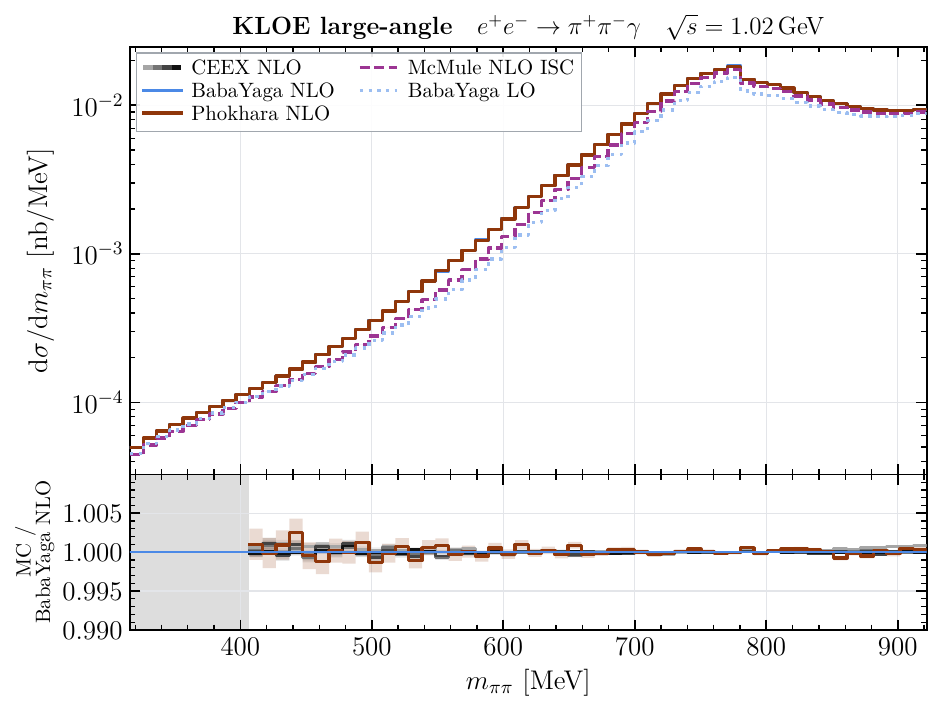}}
  \caption{Fixed-order NLO predictions for the invariant mass of the final-state charged pair in
  $e^+e^-\to\mu^+\mu^-\gamma$ with the KLOE small-angle selection (left, without VP) and in
  $e^+e^-\to\pi^+\pi^-\gamma$ with the KLOE large-angle selection (right), at
  $\sqrt s=1.02$~GeV, compared with \BabaYaga, \Phokhara and \McMule at NLO.}
  \label{fig:nlo}
\end{figure}

The fixed-order NLO predictions are obtained from the resummed expansion truncated at $\nmax=1$ for the
non-radiative and $\nmax=2$ for the radiative processes, with the amplitudes and the IR
subtraction of \cref{subsec:amplitudes}. We compare these results
against the independent calculations from \BabaYaga, \Phokhara, and \McMule, with the same
inputs and cuts and without VP.\footnote{The \Phokhara prediction is that of
the forthcoming version of the code, in which the bugs affecting the NLO corrections to
$e^+e^-\to\pi^+\pi^-\gamma$ in the large-angle selection, responsible for the discrepancies
with \BabaYaga reported in ref.~\cite{Budassi:2026lmr}, have been found and fixed during the completion of this work. This version
will be made publicly available~\cite{PhokharaWeb}.} \Cref{tab:xs-nlo}
collects the total cross-sections of the four codes for all scenarios and channels, which agree
within the statistical uncertainties wherever the same corrections are included. \Cref{fig:nlo}
shows the invariant mass of the final-state pair for
$e^+e^-\to\mu^+\mu^-\gamma$ with the KLOE small-angle selection and for $e^+e^-\to\pi^+\pi^-\gamma$
with the KLOE large-angle selection, the scan process $e^+e^-\to\pi^+\pi^-$ having been compared with
\BabaYaga in \cref{subsec:emin}.
\begin{table}[!htb]
  \centering
  \footnotesize
  \setlength{\tabcolsep}{4pt}
  \begin{tabular}{@{}llrrrr@{}}
    \toprule
    scenario & process & this work & \BabaYaga & \Phokhara & \McMule \\
    & & NLO & NLO & NLO & NLO \\
    \midrule
    CMD & $\mu^+\mu^-$ & 77.16649(21) & 77.16680(16) & -- & 77.166870(32) \\
     & $\pi^+\pi^-$ & 567.3615(11) & 567.3568(16) & -- & 566.37237(16) \\
    \midrule
    KLOE large-angle & $\mu^+\mu^-\gamma$ & 0.9924902(61) & 0.9924949(62) & 0.992435(41) & 0.9924919(65) \\
     & $\pi^+\pi^-\gamma$ & 3.227373(36) & 3.227456(31) & 3.227544(89) & 2.959070(12) \\
    \midrule
    KLOE small-angle & $\mu^+\mu^-\gamma$ & 3.487147(14) & 3.487132(14) & 3.48720(12) & 3.487152(29) \\
     & $\pi^+\pi^-\gamma$ & 16.05509(13) & 16.05672(12) & 16.05684(43) & 16.13639(11) \\
    \midrule
    BES\,III & $\mu^+\mu^-\gamma$ & 0.2642737(22) & 0.2642737(21) & 0.26407(11) & 0.2642710(30) \\
     & $\pi^+\pi^-\gamma$ & 0.0838026(11) & 0.0838058(13) & 0.0837977(34) & 0.0833695(18) \\
    \midrule
    B factories & $\mu^+\mu^-\gamma$ & 0.01617881(11) & 0.01617912(11) & 0.016228(21) & 0.01617830(40) \\
     & $\pi^+\pi^-\gamma$ & 0.010082160(87) & 0.010082200(90) & 0.01010035(88) & 0.01001464(58) \\
    \bottomrule
  \end{tabular}
  \caption{Total cross-sections in nb at fixed-order NLO inside the cuts of each scenario, including the dressed form factor in the pion channels but without VP in the muon channels, compared with \BabaYaga, \Phokhara and \McMule, the latter with the ISC only for the pions.}
  \label{tab:xs-nlo}
\end{table}

\subsection{Relevance of initial--final interference}
\label{subsec:interference}

\begin{figure}[t]
  \centering
  \includegraphics[width=\textwidth]{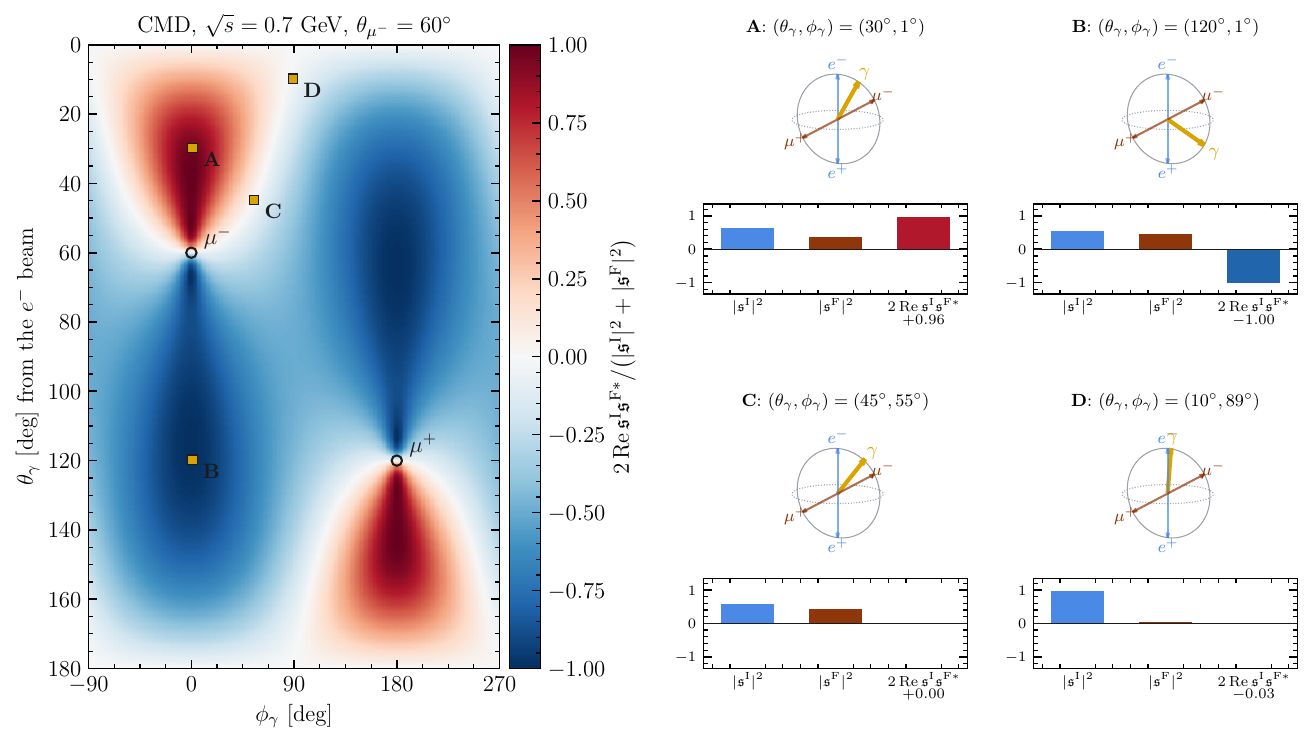}
  \caption{
  Plots showing the relative contribution of the initial--final interference of a soft photon in
  $e^+e^-\to\mu^+\mu^-\gamma$. \textbf{Left:} A map of the photonic angular phase-space showing the
  ratio of the interference term to the individual squared terms,
  $2\,{\rm Re}\,\eik^{\rm I}\eik^{{\rm F}*}/(\abs{\eik^{\rm I}}^2+\abs{\eik^{\rm F}}^2)$.
  \textbf{Right:} Plots showing the relative contributions of the $\abs{\eik^{\rm I}}^2$, $\abs{\eik^{\rm F}}^2$,
  and $2\,{\rm Re}\,\eik^{\rm I}\eik^{{\rm F}*}$ terms for four different configurations A--D,
  where the two eikonal factors are comparable and parallel (A),
  antiparallel (B), orthogonal (C), or where one of them dominates (D).
  }
  \label{fig:interference}
\end{figure}

As described in \cref{sec:methods}, the CEEX method includes the initial--final interference 
corrections to the emission of soft photons to all orders. By contrast, in the exclusive exponentiation (EEX) formalism~\cite{Jadach:1999vf,Jadach:2000ir}, implemented for
instance in \Sherpa~\cite{Krauss:2022ajk}, the factorisation of real soft divergences is performed at the level of
the amplitude squared and thus interference terms $\eik^{\rm I}_\sigma \eik^{{\rm F}*}_\sigma$
are neglected and their contributions are reinstated order by order in the finite remainders.
These interferences are large in some regions of phase-space, as
\cref{fig:interference} illustrates. Note, however, that \cref{fig:interference} does not show
the size of the interference contribution to any observable, nor does it show the size of the
eikonal interference compared with the NLO one.
In general, initial--final interference in soft-photon
radiation is the origin of the charge asymmetry of the
muon or pion pair; thus, it is important to treat this term at all orders if a precise description
of this asymmetry is to be obtained.
% The asymmetry has been measured by CMD and computed in refs.~\cite{Ignatov:2022iou,Colangelo:2022lzg}.
\section{Results}
\label{sec:results}

In this section we compare our predictions with those of the other Monte Carlo generators in the five
experimental scenarios of the RadioMonteCarLow~2 comparison~\cite{Aliberti:2024fpq}: one
in the scan mode and four in the radiative return, taking the predictions of \BabaYaga, \KKMC,
\Phokhara and \McMule from the public results of that comparison, produced with the same inputs
and cuts.\footnote{The most recent published results from \McMule, \Sherpa and \KKMC are not yet publicly available as data or code; hence, they are not included in our comparison.} All four codes
provide the LO and NLO predictions and differ in the corrections beyond NLO. \BabaYaga matches its
NLO calculation to a QED parton shower based on collinear factorisation, which resums the collinear
logarithms of the initial and final states (NLO+PS). \KKMC exponentiates the soft photons in the YFS scheme at the amplitude level, with CEEX
residuals that contain most of the NLO corrections for both the scan and the radiative-return
processes, the one-loop corrections to the radiative processes being included only partially, as
discussed in \cref{subsec:results-rad}. Its residuals also contain ingredients beyond NLO, the
leading logarithms of the two-loop vertex corrections and the exact double emission, without
reaching the complete NNLO~\cite{Jadach:2000ir}. \KKMC enters the comparison for the muon
channels only.
\McMule provides fixed-order predictions, at NNLO for the scan processes and at NLO for the
radiative-return ones, complete for the muons and restricted to the ISC for the pions. \Phokhara delivers
fixed-order NLO cross-sections, complete for both the muon and the pion channels.

Our resummed predictions are NLO+\NLLs accurate, generated with up to $\nmax=5$ photons
and a soft cut $E_{\min} = 1$~keV in the CMD and KLOE scenarios and $100$~keV at
BES\,III and the B factories. The input parameters are those of the previous
comparison, described in ref.~\cite{Aliberti:2024fpq}. In the muon channels no VP is included, neither in our predictions nor in those of the other generators, so that these are technical comparisons of the
photonic corrections rather than physical cross-sections, the VP amounting to two per mille at
CMD and about ten percent at KLOE~\cite{Ignatov:vpl}. The pion form factor is taken in the vector meson dominance
model~\cite{Sakurai:1972wk}, with the narrow $\omega$ and $\phi$ resonances parametrised as
Breit--Wigner functions and the broad $\rho$ resonances as Gounaris--Sakurai
functions~\cite{Gounaris:1968mw}. This is the harmonised parametrisation of eq.~(5.6)
and table~10 of ref.~\cite{Aliberti:2024fpq}, fitted to the measured cross-sections, which
therefore already contains the VP of the $s$-channel photon, and no further VP factor is
applied, as in \BabaYaga and \Phokhara. This form factor multiplies the point-like amplitudes outside
the loops, the F$\times$sQED approach of \cref{subsec:amplitudes} shared by all the generators of the
comparison. Since the primary motivation of this work is the study of the soft-photon
resummation, we keep this approach here. The FsQED and generalised vector-meson dominance (GVMD) approaches, in which the form factor
enters the loops through dispersion relations or vector-meson
propagators~\cite{Colangelo:2022lzg,Ignatov:2022iou,CarloniCalame:2026vfc,CarloniCalame:2026gdm,Fang:2025mhn,Fang:2026naz,PetitRosas:2026iuq},
can be included systematically in our framework. The soft photons couple to the charge of the
pions, so that the eikonal factors and the YFS form factor are unchanged, and only the amplitudes
have to be replaced, the algebraic manipulations being the sole limitation.

Each figure in this section compares our predictions with those of the other codes: the main
panel shows the distributions, the upper ratio panel the fixed-order NLO and NNLO predictions
normalised to our CEEX NLO+\NLLs prediction, and the lower ratio panel the predictions with higher
orders normalised to CEEX.
The lower ratio, higher orders/CEEX, thus compares three methods for including corrections
beyond NLO: the collinear-factorisation parton shower, the CEEX of \KKMC with its approximate
residuals, and the CEEX of this work with exact NLO residuals.
Their spread can be taken as a loose estimate of the uncertainty in the theoretical prediction,
although care should be taken in phase-space regions where a subset of the methods neglect
sizeable contributions.
The total cross-sections are collected in \cref{tab:xs}.

\begin{table}[t]
  \centering
  \footnotesize
  \setlength{\tabcolsep}{4pt}
  \begin{tabular}{@{}llrrrr@{}}
    \toprule
    scenario & process & this work & \BabaYaga & \KKMC & \McMule \\
    & & NLO+\NLLs & NLO+PS & CEEX\footnotemark & NNLO \\
    \midrule
    CMD & $\mu^+\mu^-$ & 77.21875(41) & 77.22551(46) & 77.3241(90) & 77.268239(35) \\
     & $\pi^+\pi^-$ & 569.9439(27) & 570.0422(32) & -- & 569.37990(17) \\
    \midrule
    KLOE small-angle & $\mu^+\mu^-\gamma$ & 3.486442(59) & 3.485156(34) & 3.48981(29) & -- \\
     & $\pi^+\pi^-\gamma$ & 16.02332(41) & 16.01376(20) & -- & -- \\
    \midrule
    KLOE large-angle & $\mu^+\mu^-\gamma$ & 0.980761(16) & 0.983282(10) & 0.986227(100) & -- \\
     & $\pi^+\pi^-\gamma$ & 3.192565(38) & 3.203931(43) & -- & -- \\
    \midrule
    BES\,III & $\mu^+\mu^-\gamma$ & 0.2632246(35) & 0.2640212(22) & 0.265474(102) & -- \\
     & $\pi^+\pi^-\gamma$ & 0.0834221(20) & 0.0837438(11) & -- & -- \\
    \midrule
    B factories & $\mu^+\mu^-\gamma$ & 0.01615201(40) & 0.01618183(25) & 0.0164280(28) & -- \\
     & $\pi^+\pi^-\gamma$ & 0.01002642(26) & 0.01005004(19) & -- & -- \\
    \bottomrule
  \end{tabular}
  \caption{Total cross-sections in nb inside the cuts of each scenario, including the dressed form factor in the pion channels but without VP in the muon channels, at NLO+\NLLs, compared with \BabaYaga~\cite{Balossini:2006wc,Budassi:2026lmr} matched to its parton shower (NLO+PS), \KKMC~\cite{Jadach:1999vf,Jadach:2022mbe} at CEEX and, for the scan processes, \McMule~\mbox{\cite{Banerjee:2020rww,Broggio:2022htr,Ulrich:2025fij}} at NNLO.}
  \label{tab:xs}
\end{table}
\footnotetext{The accuracy of \KKMC depends on the approximations made in its residuals $\hat\beta$, described in detail in ref.~\cite{Jadach:2000ir}: section~3.4 for the one-loop corrections to the scan processes, $\hat\beta^{(1)}_0$, and section~3.5.1 for the one-loop corrections to the single emission, $\hat\beta^{(2)}_1$, relevant to the radiative return.}

\subsection{Scan processes: CMD}
\label{subsec:results-scan}

The CMD selection at $\sqrt s=0.7$~GeV requires two charged muon or pion tracks,
satisfying
\begin{equation}
  \begin{aligned}
    &\abs{\vec p_+},\abs{\vec p_-}>0.45\,E_{\rm beam}\,,\\
    &1~{\rm rad}<\theta_{\rm avg}<\pi-1~{\rm rad}\,,\\
    &\abs{\theta_++\theta_--\pi}<0.25~{\rm rad}\,,\\
    &\abs{\abs{\phi_+-\phi_-}-\pi}<0.15~{\rm rad}\,,
  \end{aligned}
  \label{eq:cuts-cmd}
\end{equation}
where $p_\pm$ are the momenta of the $\mu^\pm$ or $\pi^\pm$,
$\phi_\pm$ are their azimuthal angles,
$\theta_\pm$ are their polar angles
with respect to the electron beam, and $\theta_{\rm avg} = \tfrac12(\theta_+ - \theta_- + \pi)$.
The last two conditions, on the acollinearity and the acoplanarity of the pair, limit the
energy of any radiated photons to a few percent of the beam energy.

\begin{figure}[t]
  \centering
  \includegraphics[width=0.49\textwidth]{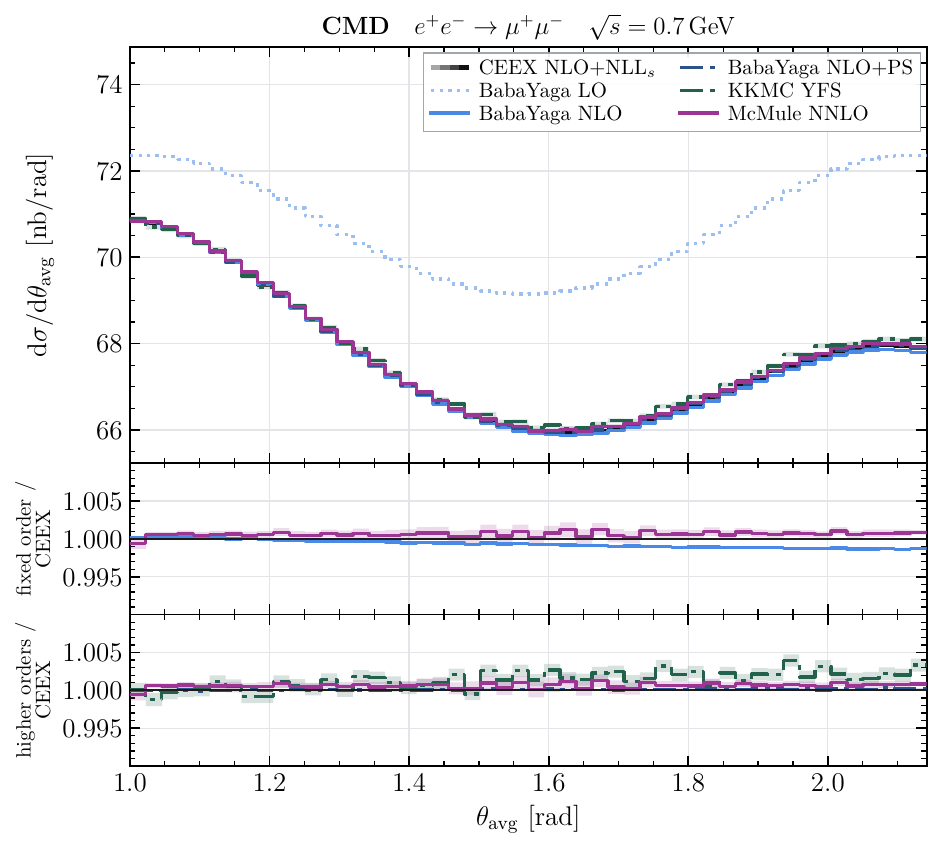}
  \includegraphics[width=0.49\textwidth]{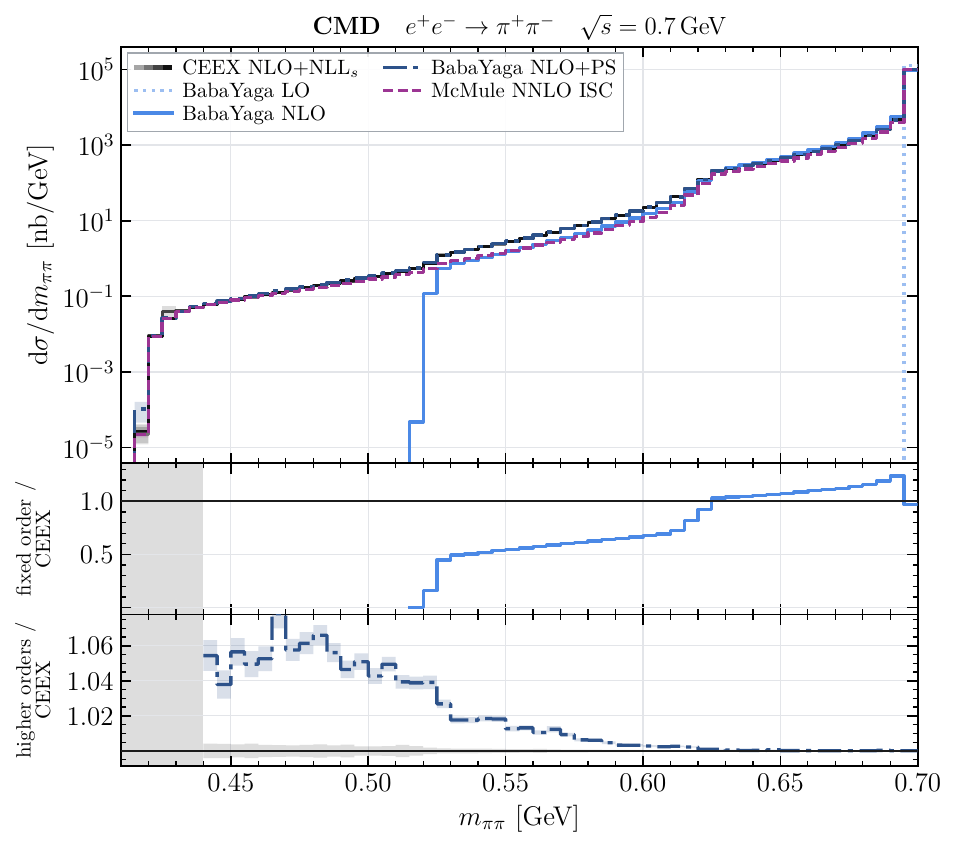}
  \caption{Average polar angle of the final-state charged pair
  for $e^+e^-\to\mu^+\mu^-$ (left, without VP) and invariant mass
  of the pair for $e^+e^-\to\pi^+\pi^-$ (right) at $\sqrt{s}=0.7$~GeV with the CMD
  selection, at NLO+\NLLs, compared with \BabaYaga, \KKMC and \McMule.}
  \label{fig:cmd3}
\end{figure}

\Cref{fig:cmd3} shows the average polar angle of the muon pair (left) and the invariant mass of
the pion pair (right). In the muon channel, we also show \BabaYaga at NLO and at NLO+PS,
\KKMC at CEEX, and \McMule at NNLO.
The total cross-section (see \cref{tab:xs}) from \BabaYaga
agrees with ours to a relative precision of $10^{-4}$, and those from \KKMC and \McMule agree at the
per-mille level.
The angular distributions agree to a few per mille, demonstrating that the different methods
for including corrections beyond NLO are compatible.
This is as expected for this scenario: the
acollinearity and acoplanarity cuts keep only soft photons, whose emission is governed by the
eikonal factors that all three schemes share;\footnote{\BabaYaga's parton shower distributes the
photon angles according to the eikonal.} the differences between a parton
shower and a soft resummation are enhanced where the photons
are hard, such as in the tail of the invariant mass. Looking at the pion invariant mass distribution
(right), we see that
below the kinematic edge at $0.5$~GeV it is
populated by hard radiation, so that the fixed-order NLO falls to half of
the resummed prediction. \BabaYaga's all-order prediction agrees with ours to
a few percent.
The \McMule prediction, which includes NNLO ISC,
is shown in the main panel for illustration.

\FloatBarrier
\subsection{Radiative return: KLOE, BES\,III and B factories}
\label{subsec:results-rad}

The radiative-return method relies on initial-state radiation to lower the effective
centre-of-mass of the collision. For these scenarios, we consider the lowest-order
processes $e^+e^-\to\mu^+\mu^-\gamma$ and $e^+e^-\to\pi^+\pi^-\gamma$.

\begin{figure}[!t]
  \centering
  \includegraphics[width=0.43\textwidth]{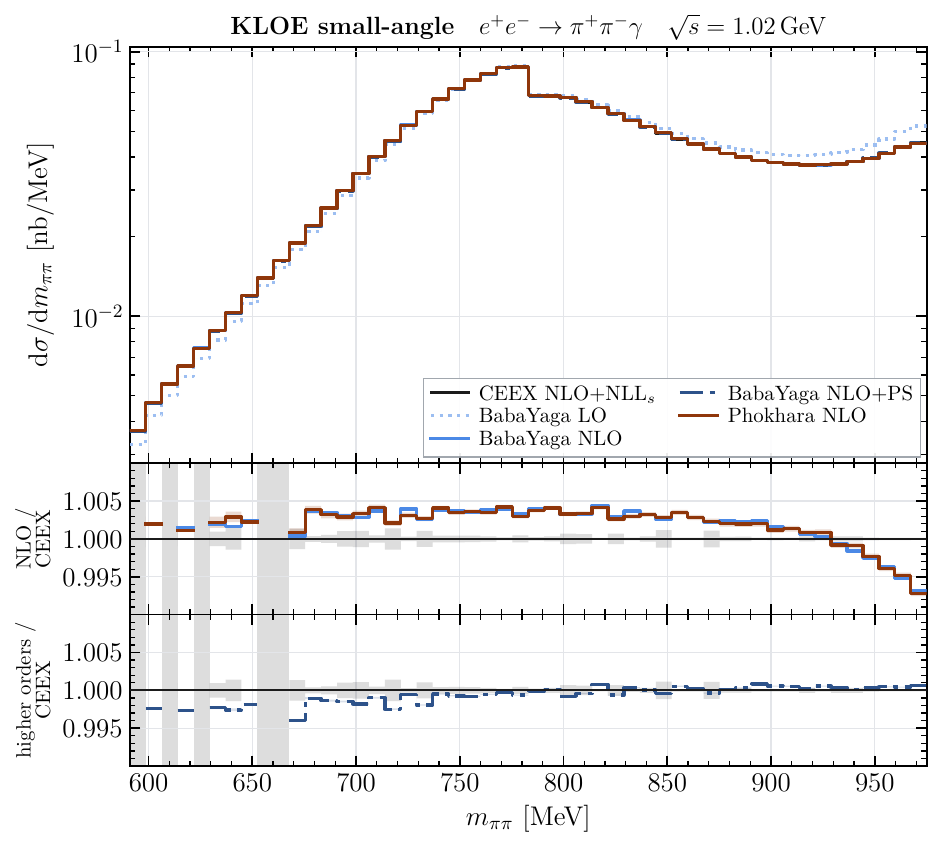}
  \includegraphics[width=0.43\textwidth]{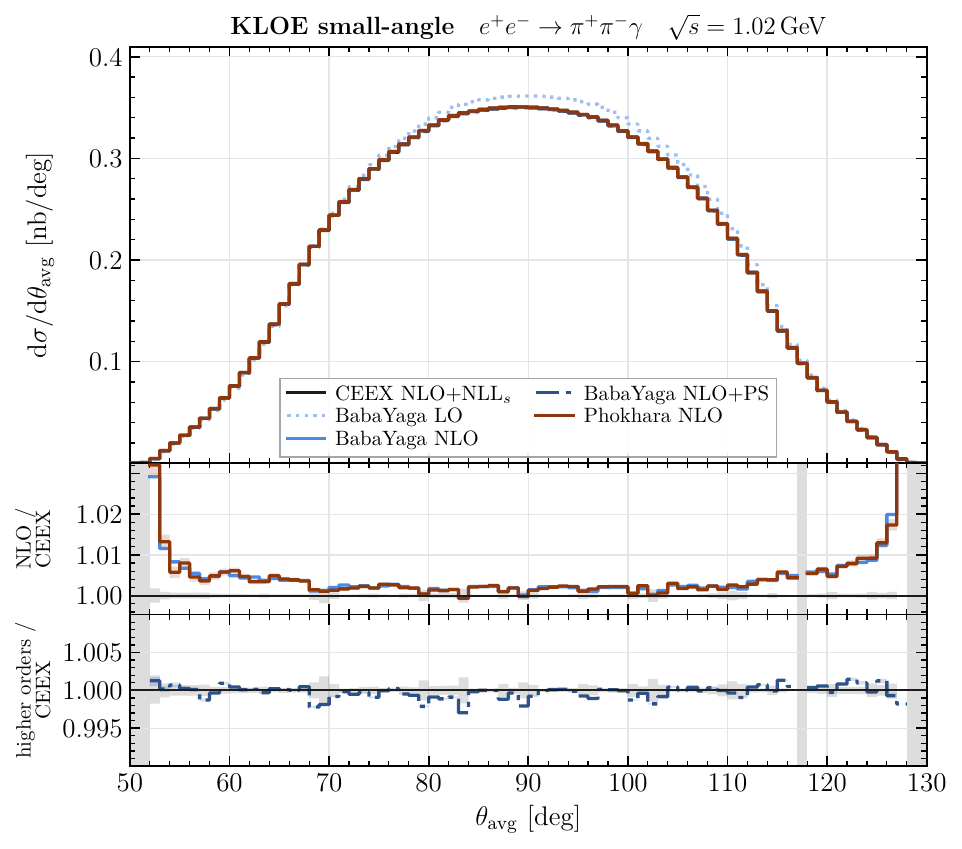}\\
  \includegraphics[width=0.43\textwidth]{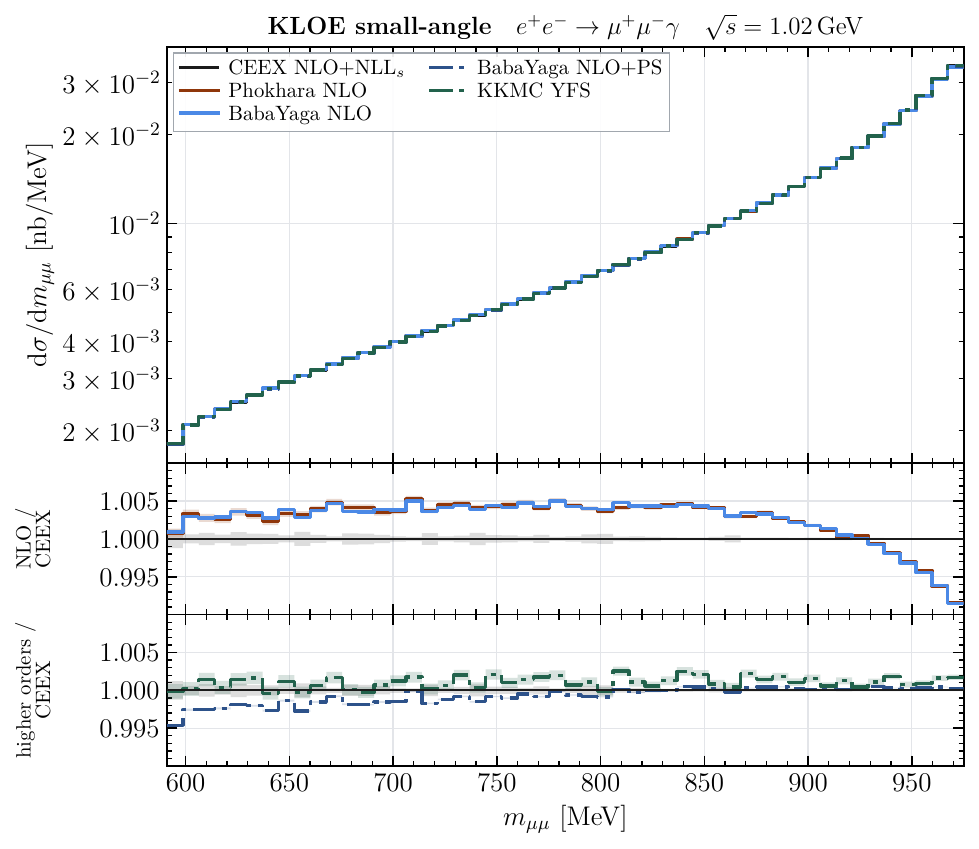}
  \includegraphics[width=0.43\textwidth]{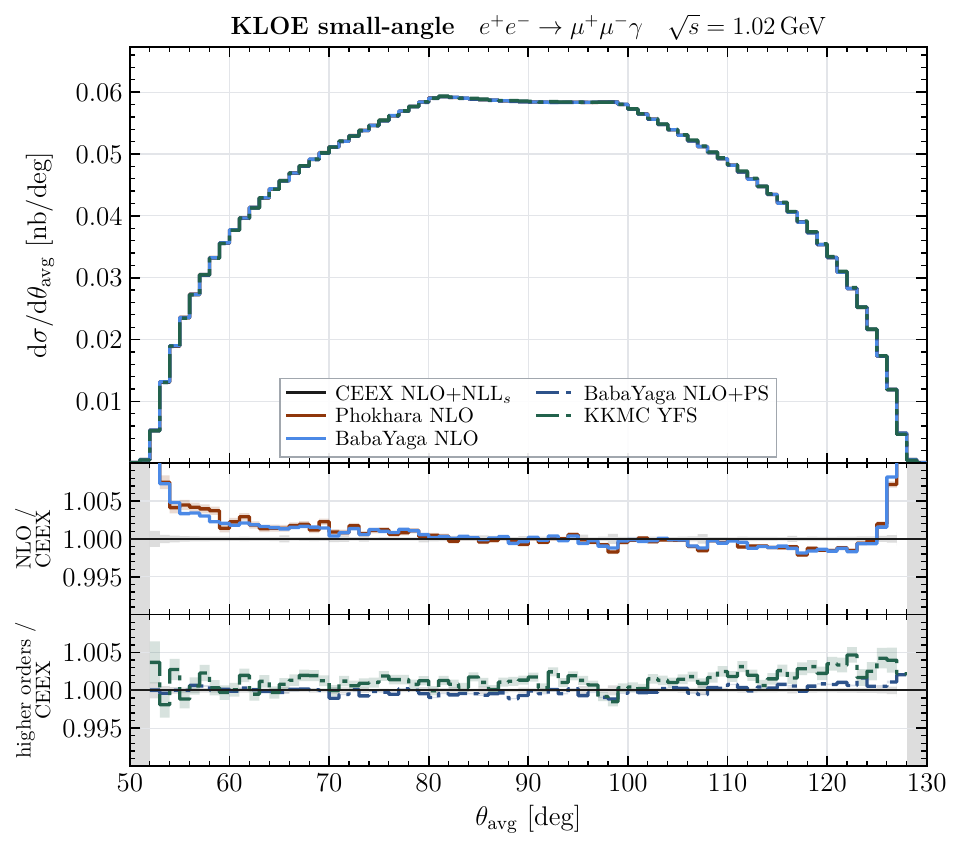}
  \caption{Invariant mass and average polar angle of the final-state charged
  pair for $e^+e^-\to\pi^+\pi^-\gamma$ (top)
  and $e^+e^-\to\mu^+\mu^-\gamma$ (bottom, without VP) at $\sqrt{s}=1.02$~GeV with the KLOE small-angle
  selection, at NLO+\NLLs, compared with \BabaYaga, \Phokhara and, for the muons, \KKMC.}
  \label{fig:kloe-sa}
\end{figure}

\paragraph{KLOE small-angle.}
The KLOE small-angle selection at $\sqrt s=1.02$~GeV requires
\begin{equation}
  \begin{aligned}
    &50^\circ<\theta_+,\theta_-<130^\circ\,,\\
    &\abs{p_{z,\pm}}>90~{\rm MeV}\ \text{or}\ p_{T,\pm}>160~{\rm MeV}\,,\\
    &\theta_{\rm miss}<15^\circ\ \text{or}\ \theta_{\rm miss}>165^\circ\,,\\
    &0.35~{\rm GeV}^2<M^2_{xx}<0.95~{\rm GeV}^2\,,
  \end{aligned}
  \label{eq:cuts-kloe-sa}
\end{equation}
where $\theta_{\rm miss}$ is the angle of the missing momentum, reconstructed from the charged final-state particles with respect to the electron beam, $p_{z,\pm}$ is the $z$ component of each charged
final-state particle's momentum, $p_{T,\pm}$ is the transverse component,
and $x \in \{\mu,\pi\}$.
The effect of these selection cuts is to enhance the contribution of collinear initial-state
radiation and suppress final-state radiation.
% This is the selection of the KLOE measurements of the pion form factor, in which the pion spectrum is
% normalised to the muon one, so that the two channels are measured and corrected in the same way.
\Cref{fig:kloe-sa} shows the invariant mass (left) and the average polar angle (right)
of the final-state pair. The top row shows the pion process
compared with \BabaYaga at NLO and NLO+PS and with \Phokhara at NLO.
The agreement is sub-per-mille between the resummed predictions in the angular distribution
and on a similar level for the majority of the invariant mass distribution. In the hard
radiation tail, our prediction lies 2 per mille higher than that of \BabaYaga.
The upper ratio plots show the ratio compared to fixed-order NLO; unlike the scan process,
the higher-order corrections for this scenario amount to less than 1\% across most of the
phase-space.
The bottom row of \cref{fig:kloe-sa} shows
the muon production process, where \KKMC is also included in the comparison. The \KKMC
prediction for the radiative-return processes includes the double-real
correction which dominates the shape of the NLO correction in kinematic distributions, as well
as the real emission process at approximate one loop.
\KKMC's invariant mass prediction agrees within a per mille with our implementation of CEEX, with
both soft resummations sitting 2 per mille above \BabaYaga's parton shower in the tail.
In the angular distribution, a small per-mille-level difference between \KKMC and our
prediction is visible; this is expected, since \KKMC includes only a subset of the one-loop
diagrams of the radiative process, the vertex-like corrections of ref.~\cite{Jadach:2000ir} with the
final-state ones in the collinear leading-logarithmic approximation, and omits the diagrams in which the virtual
photon connects the initial and final states, whereas our one-loop amplitude is complete.
For both processes, the addition of resummation gives a correction of up
to $-1\%$ in the upper end of the invariant mass, as well as roughly 0.5\% in the rest of the spectrum.

\paragraph{KLOE large-angle.}
The KLOE large-angle selection, based on the same collider energy of 1.02~GeV as the small-angle
selection, requires that a photon with energy $E_\gamma$ and polar angle $\theta_\gamma$ is detected,
such that
\begin{equation}
  \begin{aligned}
    &50^\circ<\theta_+,\theta_-<130^\circ\,,\\
    &\abs{p_{z,\pm}}>90~{\rm MeV}\ \text{or}\ p_{T,\pm}>160~{\rm MeV}\,,\\
    &50^\circ<\theta_\gamma<130^\circ\,,\qquad E_\gamma>20~{\rm MeV}\,,\\
    &0.10~{\rm GeV}^2<M^2_{xx}<0.85~{\rm GeV}^2\,,
  \end{aligned}
  \label{eq:cuts-kloe-la}
\end{equation}
where all other quantities are defined above.
The effect of these selection cuts is to amplify the hard wide-angle radiation, resulting
in an increased contribution from the FSC and initial--final interference compared with the small-angle
scenario.

\begin{figure}[!t]
  \centering
  \includegraphics[width=0.43\textwidth]{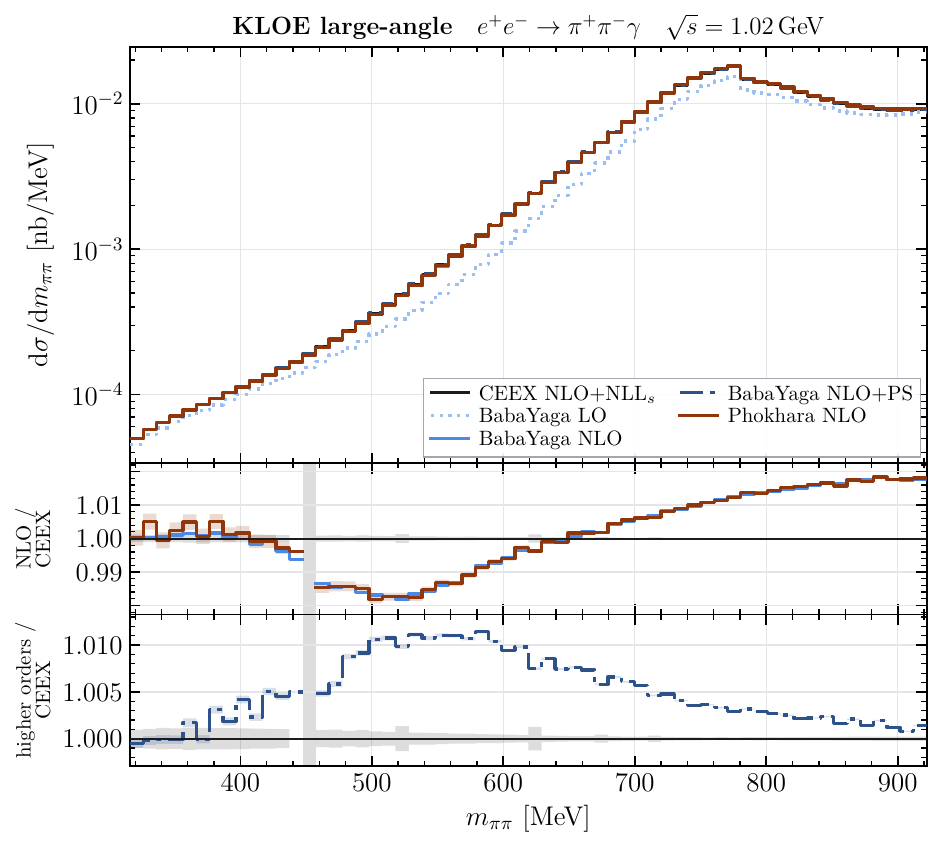}
  \includegraphics[width=0.43\textwidth]{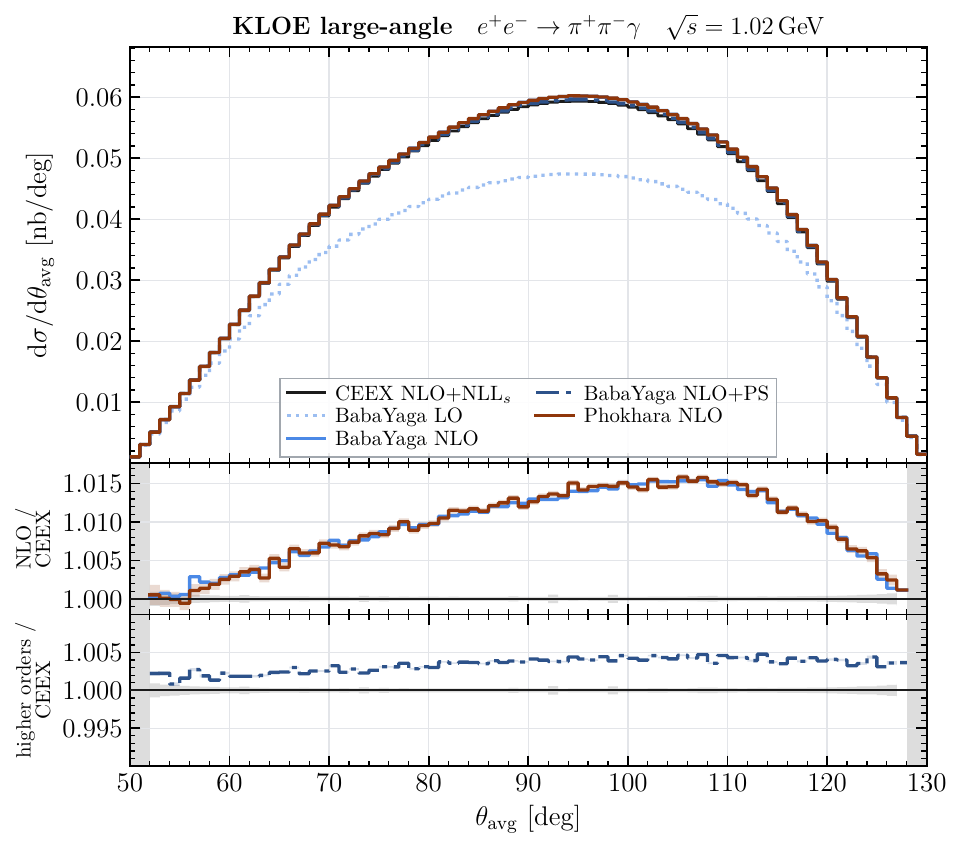}\\
  \includegraphics[width=0.43\textwidth]{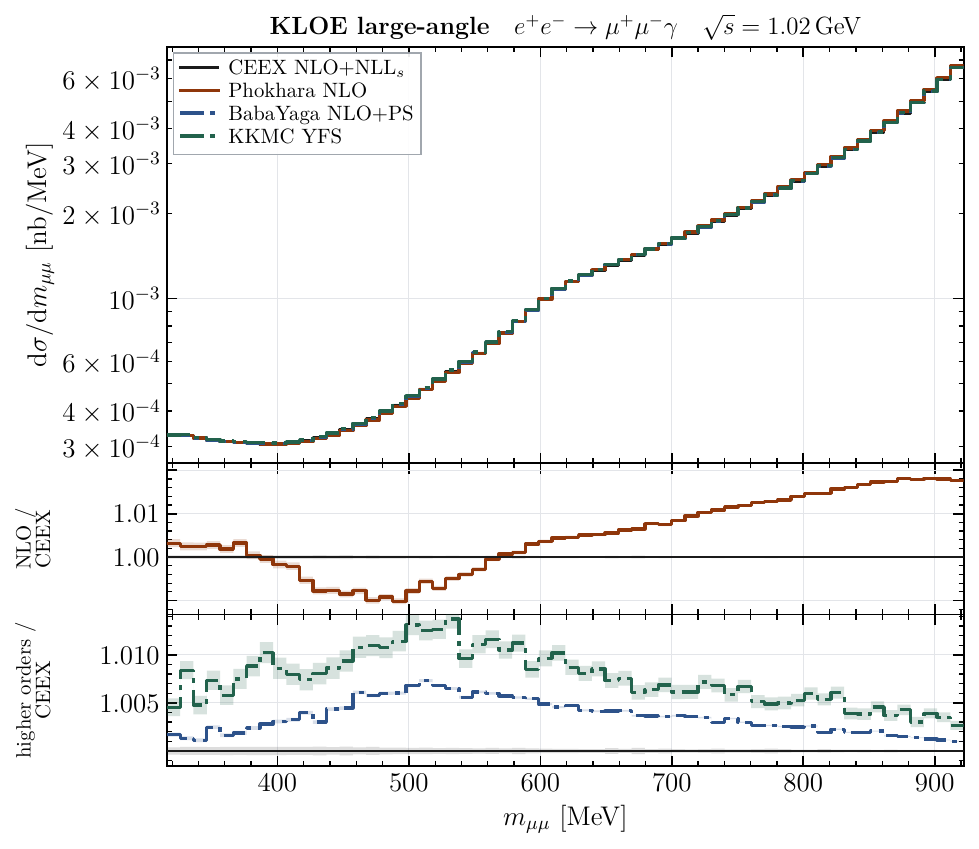}
  \includegraphics[width=0.43\textwidth]{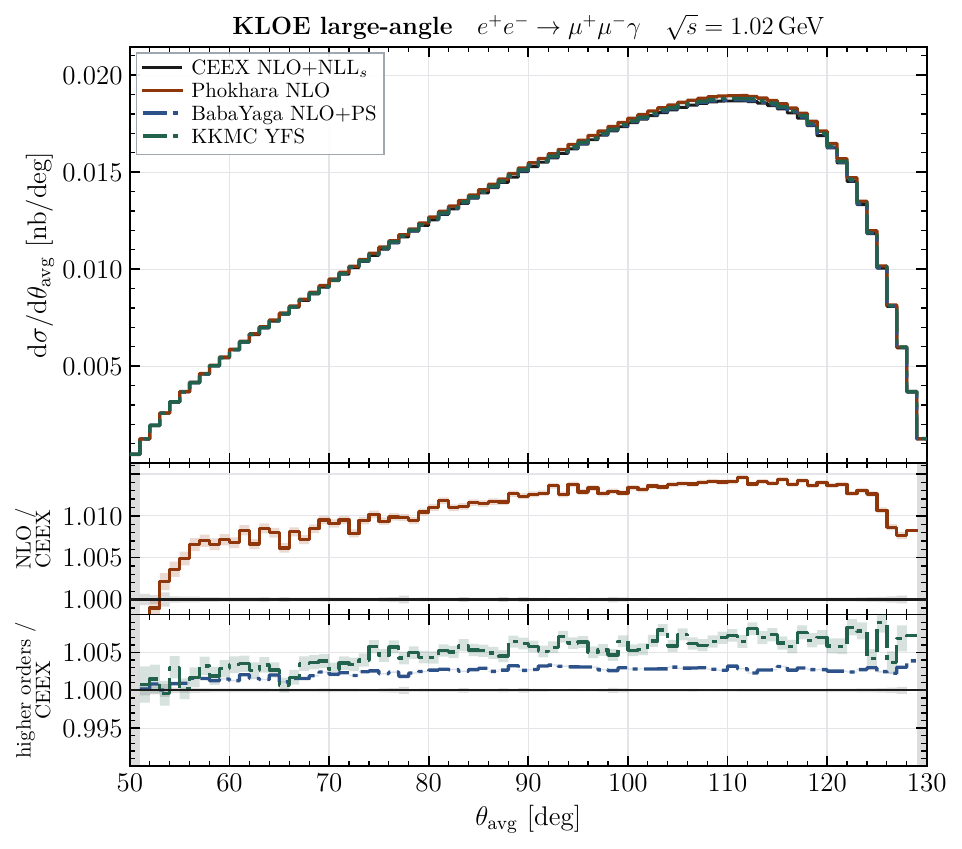}
  \caption{Invariant mass and average polar angle of the final-state charged
  pair for $e^+e^-\to\pi^+\pi^-\gamma$ (top)
  and $e^+e^-\to\mu^+\mu^-\gamma$ (bottom, without VP) at $\sqrt{s}=1.02$~GeV with the KLOE large-angle
  selection, at NLO+\NLLs, compared with \BabaYaga, \Phokhara and, for the muons, \KKMC.}
  \label{fig:kloe-la}
\end{figure}

\Cref{fig:kloe-la} compares our prediction with fixed-order and resummed predictions
from \BabaYaga, \Phokhara, and \KKMC for pion production (top) and muon production (bottom)
for the KLOE large-angle scenario. The invariant mass predictions (left) show a similar shape
of corrections beyond NLO, with corrections of up to 2\% for the pion process and 1\% for the muons.
The average polar angle (right) similarly shows 1\% negative corrections from resummation for
both processes. Between the NLO+PS and the two CEEX predictions, the discrepancies are smaller,
but sizeable. The invariant mass differs by up to 1\% between the methods, concentrated in the
region $450\,\MeV \leq m_{xx} \leq 700\,\MeV$. For the polar angle, the differences are flatter
and mostly reflect differences in the total cross-section.
Our predicted cross-sections (\cref{tab:xs}) lie 0.3\% lower than those
of \BabaYaga, and the muon cross-section lies 0.6\% lower than the \KKMC prediction.
The difference with \KKMC is likely to be due to the fact that we compute the full NLO correction
which is only approximate in these \KKMC results. However, the 0.3\% difference with \BabaYaga is
likely to be a true reflection of the different resummation paradigms employed, since \BabaYaga
also contains the full NLO corrections.

\FloatBarrier
\paragraph{BES\,III.}
BES\,III, like KLOE, is a radiative-return experiment, but the collider energy $\sqrt s=4$~GeV
is far above the $\rho$ resonance, thus the phenomenology is distinct. The selection 
for this scenario requires
\begin{equation}
  \abs{\cos\theta_\pm}<0.93\,,\qquad p_{T,\pm}>0.3~{\rm GeV}\,,
  \label{eq:cuts-bes3}
\end{equation} 
and we require exactly one detected photon to be hard, $E_\gamma>0.4$~GeV, with
detected photons satisfying
\begin{equation}
  \begin{aligned}
    \abs{\cos\theta_\gamma}<0.8 \quad&\text{with}\quad E_\gamma>25\,\MeV\\
    \text{or} \qquad 0.86<\abs{\cos\theta_\gamma}<0.92 \quad&\text{with}\quad E_\gamma>50\,\MeV\,,
  \end{aligned}
  \label{eq:cuts-bes3-2}
\end{equation}%
corresponding to the barrel and the endcaps of the detector, respectively.

\begin{figure}[!t]
  \centering
  \includegraphics[width=0.49\textwidth]{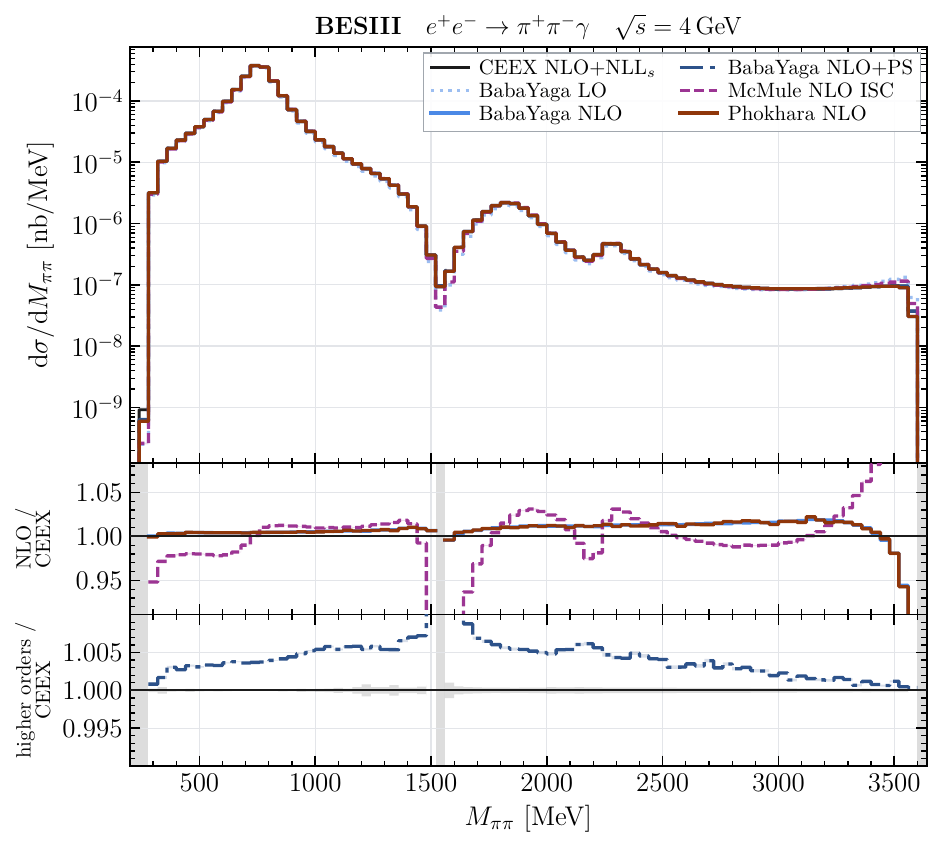}
  \includegraphics[width=0.49\textwidth]{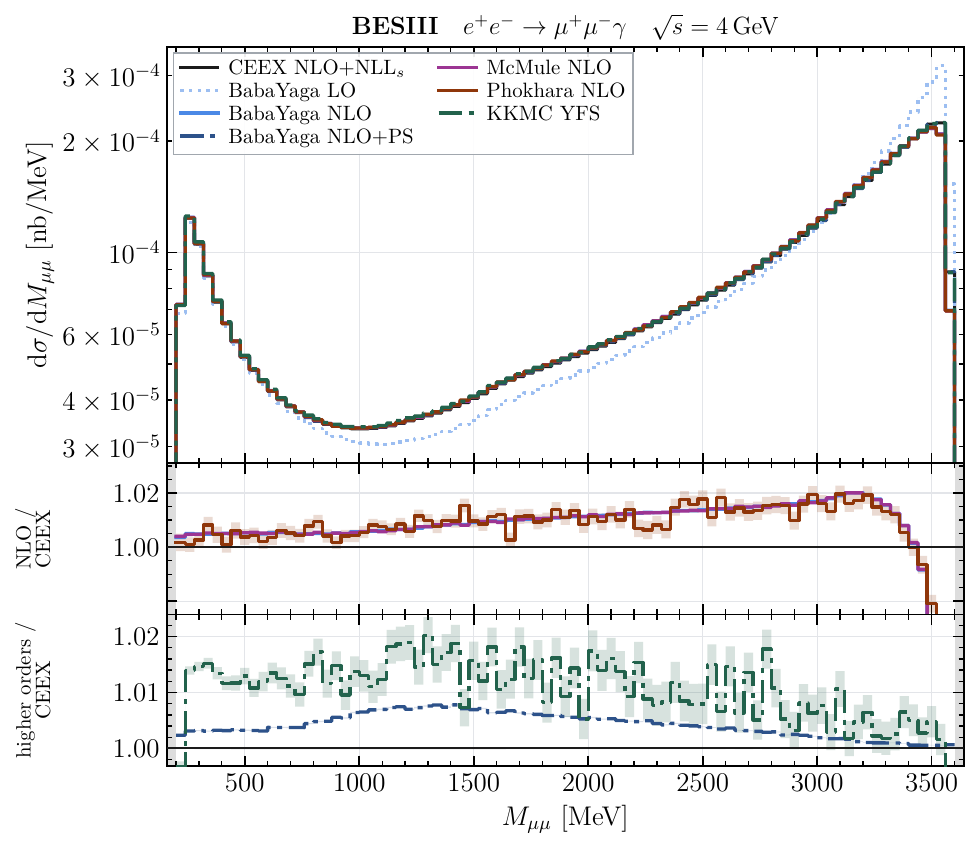}
  \caption{Invariant mass of the final-state charged pair for $e^+e^-\to\pi^+\pi^-\gamma$ (left) and
  $e^+e^-\to\mu^+\mu^-\gamma$ (right, without VP) at $\sqrt s=4$~GeV with the BES\,III selection, at NLO+\NLLs,
  compared with \BabaYaga, \McMule and, for the muons, \KKMC and \Phokhara.}
  \label{fig:bes3}
\end{figure}

\Cref{fig:bes3} shows the invariant mass distribution for the pion production process (left),
compared with \McMule NLO ISC, \BabaYaga NLO and NLO+PS, and the muon process (right) compared with
\McMule full NLO, \BabaYaga NLO and NLO+PS, \Phokhara NLO, and \KKMC at CEEX (see discussion above).
All fixed-order full NLO predictions agree at the per-mille level, whilst the correction from resummation
amounts to up to 2\% in the invariant mass region below 3200 MeV for both muons and pions. For very high
$m_{xx}$ at the limit of the selection, the corrections from radiation beyond NLO are very large. This is
due to the effect of multiple semi-soft radiation and illustrates the importance of including these corrections
in Monte Carlo generators.
Comparing the predictions of \BabaYaga NLO+PS and \KKMC{} CEEX with our result,
the cross-sections agree to a few per mille, while in the lower ratio plot,
we see that the difference is sub-percent across most of the $m_{\pi\pi}$ distribution. The exception
is in the region around 1500 MeV where the $\rho(770)$ and the $\rho(1450)$ of the pion form
factor interfere destructively and the cross-section has a dip, so that
large relative differences are seen, though the value of the cross-section is small.
For the muon process, the \BabaYaga and \KKMC predictions lie 0.3--0.9\% higher than ours.

\paragraph{B factories.}
The B-factory scenario focuses on radiative return at $\sqrt s=10$~GeV. The selection requires
a hard wide-angle photon which accounts for most of the missing energy
\begin{equation}
  \begin{aligned}
    &0.65~{\rm rad}<\theta_+,\theta_-<2.75~{\rm rad}\,,\qquad \abs{\vec p_+},\abs{\vec p_-}>1~{\rm GeV}\,,\\
    &0.6~{\rm rad}<\theta_\gamma<2.7~{\rm rad}\,,\qquad E_\gamma>3~{\rm GeV}\,,\\
    &\angle(\vec p_\gamma,\vec p_{\rm miss})<0.3~{\rm rad}\,,\\
    &M_{xx\gamma}>8~{\rm GeV}\,,
  \end{aligned}
  \label{eq:cuts-b}
\end{equation}
where $x \in \{\mu,\pi\}$ as before. These selection cuts ensure that hard collinear
initial-state (undetected) radiation is suppressed.

\begin{figure}[!t]
  \centering
  \includegraphics[width=0.49\textwidth]{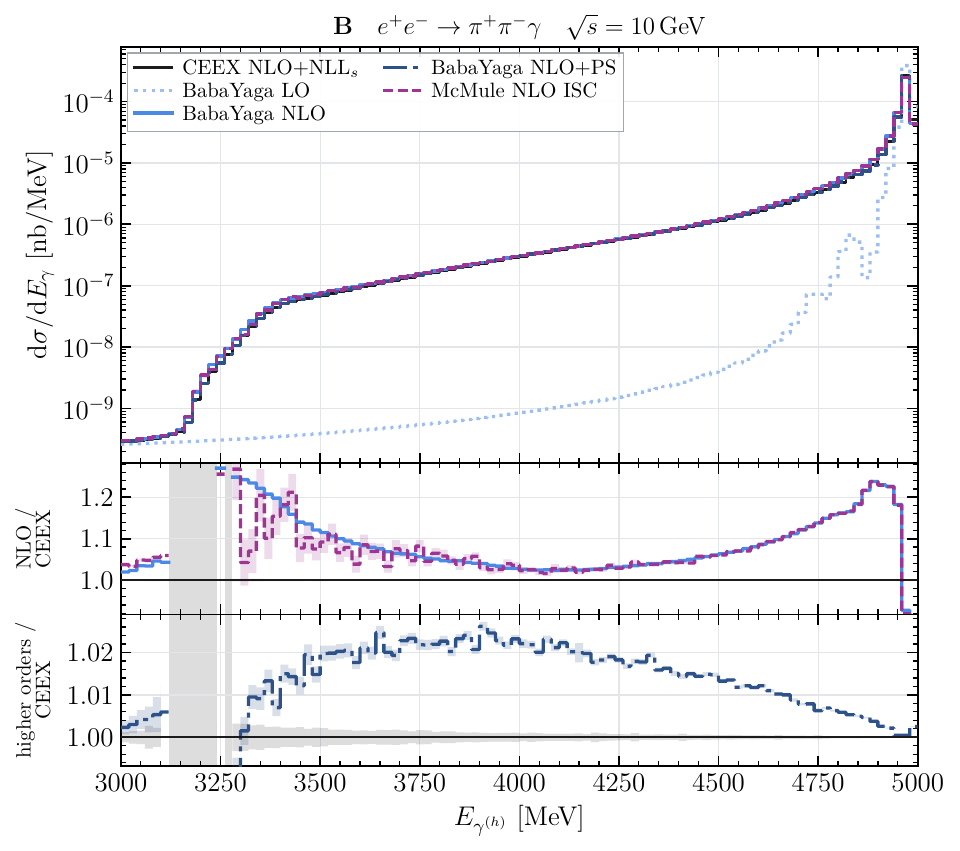}
  \includegraphics[width=0.49\textwidth]{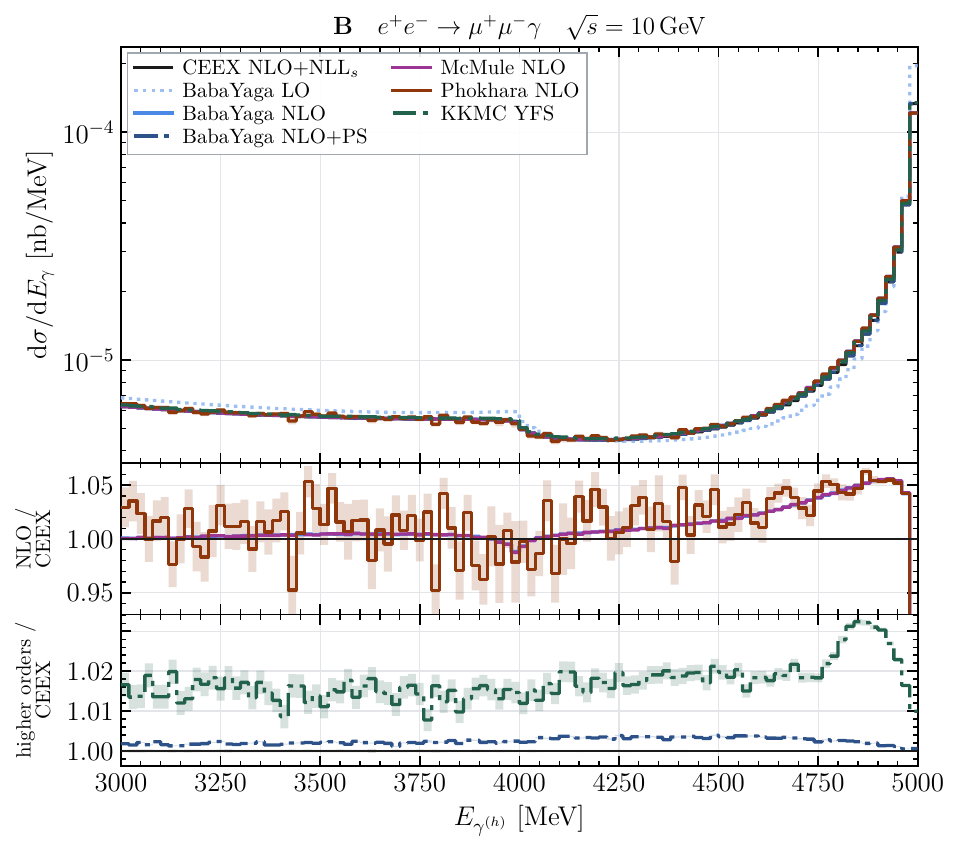}
  \caption{Photon energy for $e^+e^-\to\pi^+\pi^-\gamma$ (left) and $e^+e^-\to\mu^+\mu^-\gamma$
  (right, without VP) at $\sqrt s=10$~GeV with the B-factory selection, at NLO+\NLLs, compared
  with \BabaYaga, \McMule and, for the muons, \KKMC.}
  \label{fig:b}
\end{figure}

In this scenario, we focus on the distribution of the energy of the hardest photon, to analyse
how this observable is described by the various codes and methods. We see from the main panel of
\cref{fig:b} that the NLO corrections to this observable are large, as expected,
and much larger for the pions than for the muons. This difference occurs since at LO
the photon energy fixes the pair
mass, $m_{xx}^2=s-2\sqrt s\,E_\gamma$, so that the pion spectrum follows the pion form factor and is
suppressed away from the $\rho$ peak, whereas a second photon decouples the energy of the hardest
photon from the pair mass and fills the low-energy region with events at the $\rho$.
The form factor suppression is absent from the muon process, hence the NLO correction is
around 20\% differentially in $E_\gamma$.
The higher-order corrections beyond NLO are also sizeable, up to 20\% in the pion process, due to the sensitivity of
this observable to multiple photon radiation.
Comparing the total cross-sections of the resummed predictions, this work and \BabaYaga agree
at 2 per mille, whilst the \KKMC result is 1.7\% higher. Differentially in the hardest photon
energy, the agreement is around 1\% for pions and even better for muons,
which represents excellent convergence of the methods in describing an exclusive observable.

\FloatBarrier

Across the five scenarios, we observe excellent agreement between the radiative corrections
emerging from our resummation procedure and those of the other Monte Carlo generators. Where
the selection favours soft or collinear radiation, for inclusive observables
in the scan mode and in the KLOE
small-angle selection, the corrections beyond NLO are at the per-mille level and the three
all-order predictions agree at that level, so that the choice of the resummation procedure does
not impact the results.
For more exclusive observables, such as the hard-radiation tail of the invariant mass distribution
or for scenarios where a hard photon is tagged at large angle, the corrections beyond
NLO reach several percent in the distributions,
and the resummed predictions spread over a few tenths of a percent.
In the latter case, the theory uncertainty should be taken at the 0.1--1\% level
(depending on the scenario) only if corrections beyond NLO
have been taken into account. The sizeable differences with respect to NLO imply
that fixed-order NLO is insufficient for precise predictions of these observables at the percent level.

\section{Conclusions}
\label{sec:conclusions}

In this work, we have presented a new implementation of soft-photon resummation in the CEEX scheme for the processes $e^+e^-\to\mu^+\mu^-(\gamma)$ and $\pi^+\pi^-(\gamma)$, with all UV and IR divergences regulated dimensionally.
Since CEEX exponentiates the soft photons at the amplitude level, the interference between the initial- and final-state corrections is retained at all orders, and the resummation is matched to exact NLO matrix elements, including all the finite masses of the fermions. On top of this, we include a two-photon matrix element matching to improve the description of real radiation beyond the eikonal approximation, which is important in the tail of distributions sensitive to these, such as the invariant pair mass. The one-loop amplitudes keep the full mass dependence: those of the scan processes and of $e^+e^-\to\pi^+\pi^-\gamma$ are obtained analytically with the tensor decomposition of ref.~\cite{Dave:2026pvq}, while that of $e^+e^-\to\mu^+\mu^-\gamma$ is evaluated numerically diagram by diagram. The YFS form factor is computed with the same regulator, so that its IR poles cancel against those of the one-loop amplitudes at every phase-space point, and with a Lorentz-invariant soft-photon cut.

We have compared our predictions with other Monte Carlo tools --- \BabaYaga, \KKMC, \Phokhara, and \McMule\ --- in the energy-scan scenario of CMD and in the radiative return of KLOE, BES\,III and the B factories. Wherever the selection cuts favour soft or collinear radiation, \ie in the energy-scan mode and in the KLOE small-angle selection, all the resummed predictions agree well within a per mille in the cross-sections and to a few per mille in differential distributions.
Where a hard photon is tagged at large angle, our prediction lies $0.2$ to $1$ percent below the NLO-matched parton shower of \BabaYaga, and $0.5$ to $1.7$ percent below the CEEX prediction of \KKMC.
The fixed-order NLO differs from all resummed predictions by up to twenty percent in the distributions sensitive to further emissions.
We thus stress the importance of carefully evaluating theoretical uncertainties by comparing independent methods and implementations.

The framework presented here will be included in the upcoming public release of the \Phokhara generator~\cite{PhokharaWeb}, which will then provide \NLLs accuracy, via the soft-photon resummation, on top of its NLO predictions for the radiative-return processes $e^+e^-\to\mu^+\mu^-\gamma$ and $\pi^+\pi^-\gamma$. Work is in progress to include the ISC at NNLO in \Phokhara for the same processes~\cite{PetitRosas:2025xhm,Pozzoli:2026eiu}, which can then be matched to the resummation presented here. The inclusion of VP effects in the muon channel, the dressing of the photon propagators with the leptonic and hadronic corrections~\cite{Jegerlehner:2017gek,Jegerlehner:vpl,Keshavarzi:2019abf,Ignatov:vpl}, essential for obtaining precise physical predictions, is planned in the next \Phokhara release.
For the muon channel, the NNLO ISC have recently become available in \McMule~\cite{Fang:2026dbs} for the KLOE large-angle, BES\,III, and B-factory scenarios, which share the feature that a photon is tagged at large angle. The two-loop amplitude is known only for a massless electron, and its electron-mass dependence is recovered through massification~\cite{Penin:2005eh,Becher:2007cu,Engel:2018fsb}, up to terms suppressed by powers of $m_e^2/Q^2$, with $Q$ any other scale of the process. This approximation is therefore not justified when the photon is emitted collinear to the beams, where $Q^2\sim m_e^2$, as in the KLOE small-angle scenario, which requires full mass dependence. In addition, the FSC and the mixed initial--final corrections at NNLO must be included for a complete description of this process.
In other future work, the collinearly enhanced modification of the YFS exponentiation proposed in ref.~\cite{Jadach:2023pka}, in which the collinear logarithms of the initial-state emission are exponentiated together with the soft ones instead of being recovered order by order through the residuals, is a natural extension of the scheme presented here.
A comparison with the NNLO ISC of \McMule and with the very recent extension of \KKMC and \Sherpa to the pion channel, at NLO matched to the amplitude-level resummation~\cite{Delorme:2026qbm}, is left to future work.

\subsection*{Acknowledgements}

We are indebted to Pau Petit Ros\`as, Thomas Teubner and Graziano Venanzoni for numerous discussions at different stages of the project.
We are also grateful to the members of the KLOE-2 collaboration, as well as to Alan Price and Marek Sch\"onherr, for their input and the discussions we have had.
We thank Yannick Ulrich for discussions on the notation and the proper counting of logarithms in soft-photon resummation.
Finally, we thank the members of the RadioMonteCarLow~2 effort, whose stimulating interactions provided the motivation for this project.
Parts of the code development and the production of plotting scripts were carried out with the assistance of AI agents,
specifically Claude Opus 5.
This work is supported by the Leverhulme Trust, LIP-2021-014, and through the Ramón y Cajal contract RYC2024-050272-I, funded by MICIU/AEI/10.13039/501100011033
and FSE+.

\appendix
\crefalias{section}{appendix}\crefalias{subsection}{subappendix}
% appendices here
\section{Calculation of the YFS form factor}
\label{app:yfs}

This appendix gives the two pieces of \cref{eq:Yij-def} in dimensional regularisation
and checks against existing implementations. The six dipoles fall into two classes.
For the two beams and for the final pair, $\theta_i\theta_j=+1$, the virtual exchange is a
production process evaluated at $s_{ij}=(p_i+p_j)^2>0$. For the four initial--final pairs,
$\theta_i\theta_j=-1$, it is a scattering process evaluated at
$t_{ij}=(p_i-p_j)^2<0$. The real part does not distinguish the two cases, since the eikonal
current $p/(k\cdot p)$ is even in $p$. We first derive the form factor completely for a
production pair, then list what changes for a scattering pair.

\subsection{Production region}
\label{app:production}

\subsubsection*{Kinematics}
A dipole is characterised by the two masses $m_i$, $m_j$ and by one invariant, the scalar
product $P_{ij}=p_i\cdot p_j>0$, from which the production and the scattering invariants
are $s_{ij}=m_i^2+m_j^2+2P_{ij}$ and $t_{ij}=m_i^2+m_j^2-2P_{ij}$. The soft integrals
depend on $P_{ij}$ through the relative velocity of the two legs and through their cusp
angle
\begin{equation}
  \rho_{ij}=\frac{\sqrt{P_{ij}^2-m_i^2m_j^2}}{P_{ij}},\qquad
  L_{ij}=\ln\frac{P_{ij}(1+\rho_{ij})}{m_im_j}=\mathrm{arccosh}\frac{P_{ij}}{m_im_j}\,,
  \label{eq:pairkin}
\end{equation}
so that $\rho_{ij}=\tanh L_{ij}$. The one-loop triangle is most naturally a function of the
cusp variable
\begin{equation}
  x_{ij}=e^{-L_{ij}}=\frac{m_im_j}{P_{ij}(1+\rho_{ij})}\in(0,1)\,,
  \label{eq:xij}
\end{equation}
the root inside the unit circle of $x+1/x=2P_{ij}/(m_im_j)$. In the scattering region the
triangle is evaluated at $x_{ij}$ itself, in the production region at $-x_{ij}+i0$, which
is the origin of the imaginary parts below. The $\epsilon$ expansion of the masters is
most compact in terms of the two roots $y_1>y_2$ of
$s_{ij}y^2-(s_{ij}+m_i^2-m_j^2)y+m_i^2=0$
\begin{equation}
  y_{1,2}=\frac{s_{ij}+m_i^2-m_j^2\pm\lambda^{1/2}(s_{ij},m_i^2,m_j^2)}{2s_{ij}}\,,
  \label{eq:y12}
\end{equation}
which satisfy $m_i^2=s_{ij}y_1y_2$, $m_j^2=s_{ij}(1-y_1)(1-y_2)$, $0<y_2<y_1<1$ and
$x_{ij}=m_im_j/(s_{ij}y_1(1-y_2))$.

\subsubsection*{Master integrals}
With $p=p_i$ and $q=p_j$ both incoming, $(p+q)^2=s_{ij}$, the propagators of the one-loop
exchange are $D_1=(k-p)^2-m_i^2$, $D_3=(k+q)^2-m_j^2$ and the photon $D_2=k^2$, and the
integrand of $B_{ij}$ in \cref{eq:BBtilde} expands into
\begin{equation}
  \frac{4m_i^2}{D_2D_1^2}+\frac{4m_j^2}{D_2D_3^2}+\frac{8P_{ij}}{D_2D_1D_3}
  -\frac{1}{D_1^2}-\frac{1}{D_3^2}+\frac{2}{D_1D_3}\,,
  \label{eq:Bintegrand}
\end{equation}
the first three terms being the IR-divergent ones and the last three the
UV-divergent ones, whose divergences cancel. All six are members of the family
\begin{equation}
  I_{a_1a_2a_3}=e^{\gamma_E\epsilon}s_{ij}^\epsilon\int\frac{\done^Dk}{i\pi^{D/2}}\,
  \frac{1}{D_1^{a_1}D_2^{a_2}D_3^{a_3}}\,,
  \label{eq:family}
\end{equation}
which integration by parts reduces to three masters of uniform transcendental weight: the
triangle $g_1=\epsilon^2r_{ij}I_{111}$, with $r_{ij}=\lambda^{1/2}(s_{ij},m_i^2,m_j^2)=s_{ij}(y_1-y_2)$,
and the two tadpoles $g_2=\epsilon I_{002}$ and $g_3=\epsilon I_{200}$. Writing
$g_k=\sum_n\epsilon^ng_k^{(n)}$ and $L_i=\ln(s_{ij}/m_i^2)$, $L_j=\ln(s_{ij}/m_j^2)$, the
expansions in the production region are
\begin{align}
  g_1^{(0)}&=0,\qquad g_1^{(1)}=\tfrac12\ln\rbr{1-\tfrac1{y_1}}-\tfrac12\ln\rbr{1-\tfrac1{y_2}}+i\pi,
  \label{eq:g11}\\
  g_1^{(2)}&=\DiLog{\frac{y_1-y_2}{1-y_2}}-\DiLog{1-\frac{y_1}{y_2}}+\ln^2(1-y_2)-\ln^2y_2\notag\\
  &\quad+\tfrac14L_i^2-\tfrac14L_j^2-2i\pi\ln(y_1-y_2)-\pi^2,
  \label{eq:g12}\\
  g_2&=1+\epsilon L_j+\epsilon^2\rbr{\tfrac12L_j^2+\tfrac{\pi^2}{12}},\qquad
  g_3=1+\epsilon L_i+\epsilon^2\rbr{\tfrac12L_i^2+\tfrac{\pi^2}{12}}.
  \label{eq:g23}
\end{align}
The real part of $g_1^{(1)}$ is $-L_{ij}$, the cusp angle, and its imaginary part $i\pi$ is
the Coulomb phase of the production pair. Both follow from $\ln(-x_{ij}+i0)=\ln x_{ij}+i\pi$
with $\ln x_{ij}=-L_{ij}$.

\subsubsection*{Virtual part}
Reducing \cref{eq:Bintegrand} to the masters and extracting the factor
$(\mu^2/s_{ij})^\epsilon$ that fixes the scale of the soft logarithm, the virtual part takes
the form
\begin{equation}
  2\alpha\,\mathrm{Re}\,B_{ij}=\frac{\alpha}{4\pi}\,\mathrm{Re}\cbr{S_\epsilon
  \rbr{\frac{\mu^2}{s_{ij}}}^{\epsilon}\sbr{\frac{c_{ij}}{\epsilon}+v_{ij}}},\qquad
  S_\epsilon=(4\pi e^{-\gamma_E})^\epsilon,
  \label{eq:virtual}
\end{equation}
where the coefficient of the pole and the finite part are
\begin{align}
  c_{ij}&=-4\sbr{1-\frac{m_i^2+m_j^2-s_{ij}}{r_{ij}}\,g_1^{(1)}},\qquad
  \mathrm{Re}\,c_{ij}=4\rbr{\frac{L_{ij}}{\rho_{ij}}-1},
  \label{eq:cij-app}\\
  v_{ij}&=4-2(L_i+L_j)+2c_{ij}-\frac{m_i^2-m_j^2}{s_{ij}}(L_i-L_j)
  +\frac{4(m_i^2+m_j^2-s_{ij})}{r_{ij}}\,g_1^{(2)}\notag\\
  &\quad+\frac{6s_{ij}-4(m_i^2+m_j^2)-\frac2{s_{ij}}(m_i^2-m_j^2)^2}{r_{ij}}\,g_1^{(1)}.
  \label{eq:vij}
\end{align}
The pole coefficient is the soft anomalous dimension of the massive dipole: the cusp
logarithm $L_{ij}/\rho_{ij}$ comes from the exchange term of \cref{eq:Bintegrand} and
the constant $-1$ from the two self-energy-like terms. Both $c_{ij}$ and $v_{ij}$ are
complex through $g_1^{(1)}$ and $g_1^{(2)}$. Only their real parts enter $Y$, while the
complex values enter $B$ of \cref{eq:BV}.

\subsubsection*{Real part}
$\tilde B_{ij}$ is Lorentz invariant and is evaluated in the rest frame of the pair, with
$\vec p_i$ along the $z$ axis, where a photon of energy $|\vec k|$ and polar angle
$\theta$, $c=\cos\theta$, has
\begin{equation}
  k\cdot p_i=\tfrac12|\vec k|\sqrt{s_{ij}}\,(a-rc),\qquad
  k\cdot p_j=\tfrac12|\vec k|\sqrt{s_{ij}}\,(2-a+rc),\qquad a=y_1+y_2,\quad r=y_1-y_2\,.
  \label{eq:kp}
\end{equation}
The cut of \cref{eq:omega} becomes $|\vec k|\,(E-\vec P\cdot\hat k)<E_{\min}\sqrt{s_{ij}}$,
where $(E,\vec P)$ is the beam momentum $p_1+p_2$ scaled by $\sqrt{s/s_{ij}}$ and boosted
to the pair frame, and $\Theta$ denotes the angle between $\vec P$ and $\vec p_i$. The
radial integral is done in $D$ dimensions, the azimuthal one in closed form with
$\int_0^{2\pi}\done\phi\,\ln(A+B\cos\phi)=2\pi\ln\tfrac12(A+\sqrt{A^2-B^2})$, and the polar
integral is left as a one-dimensional quadrature. The result has the same structure as the
virtual part, with the scale $E_{\min}$ in place of $\sqrt{s_{ij}}$
\begin{equation}
  \begin{aligned}
  2\alpha\tilde B_{ij}(\Omega)&=\frac{\alpha}{4\pi}\,\mathrm{Re}\cbr{S_\epsilon
  \rbr{\frac{\mu^2}{E_{\min}^2}}^{\epsilon}\sbr{-\frac{c_{ij}}{\epsilon}+q_{ij}}},\\
  q_{ij}&=8r^2\int_{-1}^{1}\done c\,\frac{1-c^2}{\sbr{(a-rc)(2-a+rc)}^2}\,
  \ln\frac{2\sqrt{s_{ij}}\sqrt{1-c^2}}{A+\sqrt{A^2-B^2}}\,,
  \end{aligned}
  \label{eq:real}
\end{equation}
with $A=E-|\vec P|\cos\Theta\,c$ and $B=|\vec P|\sin\Theta\sqrt{1-c^2}$. The coefficient of
the pole is $-\mathrm{Re}\,c_{ij}$, so that the real and the virtual poles cancel pair by
pair. The quadrature is kept in the code because it is fast and free of the large
cancellations of the closed forms at electron masses. For a pair at rest in the beam frame,
the beams themselves or a final pair without initial-state emission, $\vec P=0$ and the integral has a closed
form in terms of $u=y_1/y_2$ and $v=(1-y_2)/(1-y_1)$
\begin{equation}
  \begin{aligned}
    q_{ij}={}&-8\ln2+\frac{2}{y_1-y_2}\sbr{(y_1+y_2)\ln u+(2-y_1-y_2)\ln v}\\
    &+\frac1{\rho_{ij}}\Big[4\DiLog{\tfrac1u}+4\DiLog{\tfrac1v}-\ln^2u-\ln^2v\\
    &\qquad+4\ln u\,\ln\frac{2y_1}{y_1-y_2}+4\ln v\,\ln\frac{2(1-y_2)}{y_1-y_2}-\frac{4\pi^2}{3}\Big].
  \end{aligned}
  \label{eq:qrest}
\end{equation}

\subsubsection*{Result}
Adding \cref{eq:virtual,eq:real} the poles cancel, the scale $\mu$ drops
out, and the soft logarithm is left at the scale $s_{ij}$ of the pair
\begin{equation}
  Y_\Omega(p_i,p_j)=\frac{\alpha}{4\pi}\,\mathrm{Re}\sbr{c_{ij}\ln\frac{E_{\min}^2}{s_{ij}}+q_{ij}+v_{ij}}\,,
  \label{eq:Yprod}
\end{equation}
which is \cref{eq:Yij}.

\subsection{Scattering region}
\label{app:scattering}

For an initial--final pair the virtual exchange is evaluated at $t_{ij}<0$ instead of
$s_{ij}$, which changes the production result in the following way and in no other.

The masters are the same triangle and tadpoles, now evaluated at $t_{ij}<0$, where they
are real: the triangle is a function of $+x_{ij}$ instead of $-x_{ij}+i0$, so the $i\pi$ of
\cref{eq:g11}, the Coulomb phase, disappears and the real parts of $g_1^{(1)}$ and
$g_1^{(2)}$ change. The pole coefficient is unchanged, $c^{(t)}_{ij}=\mathrm{Re}\,c_{ij}$
of \cref{eq:cij-app}, and is now real. The scattering integral comes out with its
soft logarithm at the scale $m_im_j$. We normalise it with the same $(\mu^2/s_{ij})^\epsilon$
as in \cref{eq:virtual}, so that the soft logarithm of every pair is at its $s_{ij}$,
and the finite part becomes
\begin{equation}
  v^{(t)}_{ij}=4F_t+c^{(t)}_{ij}\ln\frac{s_{ij}}{m_im_j}\,,
  \label{eq:vt}
\end{equation}
where the logarithm is the change of scale and $F_t$, the finite part at the scale $m_im_j$,
reads, with $P=P_{ij}$, $\rho=\rho_{ij}$, $L=L_{ij}$, $x=x_{ij}$ and $t=t_{ij}$
\begin{align}
  F_t&=-1+\frac{m_i^2-m_j^2}{2t}\ln\frac{m_i}{m_j}-\frac{P\rho L}{t}+\frac{G_t}{\rho},
  \label{eq:Ft}\\
  G_t&=-\tfrac12L^2-2L\ln(1-x^2)-\frac{\pi^2}{6}+\DiLog{x^2}\notag\\
  &\quad+\tfrac12\ln^2\frac{m_i}{m_j}+\DiLog{1-x\frac{m_i}{m_j}}+\DiLog{1-x\frac{m_j}{m_i}}.
  \label{eq:Gt}
\end{align}
The real part, \cref{eq:real}, is unchanged, and so is the cancellation of the poles.
The form factor of an initial--final pair is therefore \cref{eq:Yprod} with
$c_{ij}\to c^{(t)}_{ij}$ and $v_{ij}\to v^{(t)}_{ij}$, and the same replacement is made in
$B$ of \cref{eq:BV}.

\subsection{Validation against the literature}
\label{app:checks}

\subsubsection*{Translation from a photon mass to dimensional regularisation}
The results in the literature are regularised with a photon mass $\lambda$. At one loop the
two regularisations are related by the replacement $\ln\lambda^2\leftrightarrow1/\epsilon+\ln\mu^2$,
applied separately to the virtual and to the real part, so that the photon-mass expressions can
be compared piece by piece with $v_{ij}$ and $q_{ij}$.

\subsubsection*{Checks}

\begin{figure}[t]
  \centering
  \includegraphics[width=0.6\textwidth]{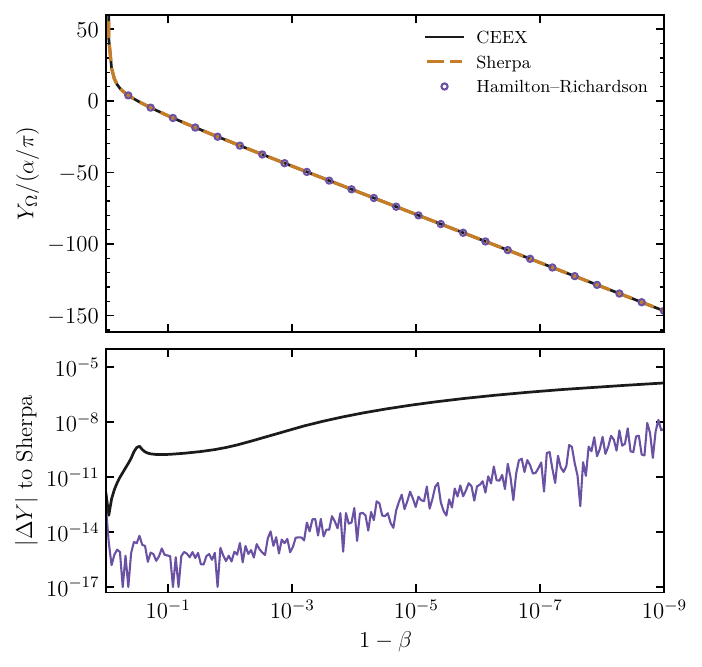}
  \caption{YFS form factor $Y_\Omega$ of an equal-mass pair at rest in the cut frame, with
  $E_{\min}=0.01\sqrt{s_{ij}}$, as a function of $1-\beta$: \cref{eq:Yprod} (this work), the
  photon-mass result of refs.~\cite{Jadach:1995sp,Krauss:2022ajk} translated to dimensional
  regularisation (\Sherpa) and the result of ref.~\cite{Hamilton:2006xz}, with the lower panel
  showing the deviation from the \Sherpa curve, $\abs{Y-Y^{\rm Sherpa}}/\max(1,\abs{Y^{\rm Sherpa}})$.}
  \label{fig:yfs-scan}
\end{figure}

The virtual part $v_{ij}$ agrees with the dimensional-regularisation expression of
ref.~\cite{Passarino:2001wy} for an equal-mass pair, up to the sign convention of the pole, and
the full form factor of \cref{eq:Yprod} has been compared with the photon-mass results of
refs.~\cite{Jadach:1995sp,Krauss:2022ajk}, valid in any frame, and with that of
ref.~\cite{Hamilton:2006xz}, valid for a pair at rest in the cut frame. In that frame the three
can be compared directly, and \cref{fig:yfs-scan} shows the agreement for an equal-mass
pair over the whole range of the velocity $\beta=\sqrt{1-4m^2/s_{ij}}$ of its two legs.

%\clearpage

\section{Tensor structures of the scan processes}
\label{app:tensors}

The tree-level ($L=0$) and one-loop ($L=1$) amplitudes of $e^+e^-\to\pi^+\pi^-$ and
$e^+e^-\to\mu^+\mu^-$ of \cref{subsec:amplitudes} are decomposed as
\begin{equation}
  \mathcal M^{(L)}=\sum_i F^{(L)}_{x,i}\,\mathcal T_{x,i}\,,\qquad x\in\{\pi,\mu\}\,,
  \label{eq:tensor-decomposition}
\end{equation}
where the tensor structures $\mathcal T_{x,i}$ form a four-dimensional basis built from the
external spinors and momenta, and the scalar form factors $F^{(L)}_{x,i}$ are obtained by
projection. For the pions, treated as point-like scalars, the basis consists of two structures,
\begin{equation}
  \mathcal T_{\pi,1}=\bar u(p_1)\,\slashed{p}_3\,v(p_2)\,,\qquad
  \mathcal T_{\pi,2}=\bar u(p_1)\,v(p_2)\,,
  \label{eq:tensors-pipi}
\end{equation}
of which only $\mathcal T_{\pi,1}$ contributes at tree level. For the muons, it consists of eight structures,
\begin{equation}
  \begin{aligned}
  \mathcal T_{\mu,1}&=\bar u(p_1)\slashed{p}_3 v(p_2)\;\bar u(p_3)v(p_4)\,,&
  \mathcal T_{\mu,2}&=\bar u(p_1)\slashed{p}_3 v(p_2)\;\bar u(p_3)\slashed{p}_1 v(p_4)\,,\\
  \mathcal T_{\mu,3}&=\bar u(p_1)\gamma^\sigma v(p_2)\;\bar u(p_3)\gamma_\sigma v(p_4)\,,&
  \mathcal T_{\mu,4}&=\bar u(p_1)\gamma^\sigma v(p_2)\;\bar u(p_3)\slashed{p}_1\gamma_\sigma v(p_4)\,,\\
  \mathcal T_{\mu,5}&=\bar u(p_1)v(p_2)\;\bar u(p_3)\slashed{p}_1\slashed{p}_2 v(p_4)\,,&
  \mathcal T_{\mu,6}&=\bar u(p_1)v(p_2)\;\bar u(p_3)v(p_4)\,,\\
  \mathcal T_{\mu,7}&=\bar u(p_1)\slashed{p}_3\gamma^\sigma v(p_2)\;\bar u(p_3)\gamma_\sigma v(p_4)\,,&
  \mathcal T_{\mu,8}&=\bar u(p_1)\slashed{p}_3 \gamma^\sigma v(p_2)\;\bar u(p_3)\slashed{p}_1 \gamma_\sigma v(p_4)\,,
  \end{aligned}
  \label{eq:tensors-mumu}
\end{equation}
of which only $\mathcal T_{\mu,3}$ contributes at tree level. The structures are evaluated on the
Kleiss--Stirling spinors of \cref{subsec:amps} for every helicity configuration, and the form
factors, reduced to the master integrals of \cref{subsec:amplitudes}, are evaluated once per
phase-space point.

\section{Phase-space channels}
\label{app:channels}

Each outgoing particle's momentum is generated from a product of energy and angular
distributions in phase-space,
where each distribution is constructed as a sum of channels.
The weight of each channel is chosen by hand for each particle flavour and also
differs for each scenario, to maximise efficiency.

The channels which are primarily used for the generation of soft-photon momenta are given
explicitly in \cref{sec:ps}, and we outline the remaining ones in \cref{tab:channels}.
The first five channels (soft, window, inverse mass, $\rho$, and spectrum) are used for the energy distribution of photons, the next four (flat, beam, cone, tag) for their angular distribution, and the last three (Born, track, final-state) for the charged final-state particles.

\begin{table}[H]
  \centering
  \small
  \begin{tabular}{@{}l>{\raggedright\arraybackslash}p{10.5cm}@{}}
    \toprule
    channel & density \\
    \midrule
    soft & $1/E$ on $[E_{\min},E_{\max}]$ \\
    window & $1/E$ over the acceptance of $M^2_{xx}$ \\
    inverse mass & $1/M^2_{xx}$ \\
    $\rho$ & Breit--Wigner of the $\rho$ in $M^2_{xx}$ \\
    spectrum & one-photon spectrum $\sigma_{\rm Born}(M^2_{xx})\,\big(1+(M^2_{xx}/s)^2\big)/(1-M^2_{xx}/s)$, over the acceptance of $M^2_{xx}$ \\
    \midrule
    flat & uniform in the solid angle \\
    tag & uniform in the angular acceptance of the tagged photon \\
    beam & $\propto 1/(1-\beta_e^2\cos^2\theta)$ about the beam axis \\
    cone & the \texttt{beam} channel restricted to a small-angle cone \\
    \midrule
    Born & $\propto 1+b\cos^2\theta^*$ in the rest frame of the final-state pair \\
    track & the Born distribution restricted to the angular acceptance of the charged tracks \\
    final-state & $\propto 1/(1-\beta^{*2}\cos^2\theta^*_k)$ about the direction of one of the photons in the rest frame of the final-state charged pair \\
    \bottomrule
  \end{tabular}
  \caption{Descriptions of the channels used for sampling the energy and the direction of the photons and for the direction of the charged final-state pair.}
  \label{tab:channels}
\end{table}

\section{Phase-space points of the pole check}
\label{app:pole-points}

\Cref{tab:pole-momenta} lists the phase-space points at which the pieces of the IR pole
of \cref{tab:pole} are evaluated, the first accepted point of the fixed-seed sample of
\cref{fig:pole} for each process.
\begin{table}[H]
  \centering
  \footnotesize
  \begin{tabular}{llrrrr}
    \toprule
    process & & $E$ & $p_x$ & $p_y$ & $p_z$ \\
    \midrule
    $\mu^+\mu^-$ & $p_+$ & $0.3500000000$ & $-0.2643092743$ & $\phantom{-}0.1791681546$ & $\phantom{-}0.0968281342$ \\
                 & $p_-$ & $0.3500000000$ & $\phantom{-}0.2643092743$ & $-0.1791681546$ & $-0.0968281342$ \\
    $\pi^+\pi^-$ & $p_+$ & $0.3500000000$ & $-0.2542466394$ & $\phantom{-}0.1723469648$ & $\phantom{-}0.0931417476$ \\
                 & $p_-$ & $0.3500000000$ & $\phantom{-}0.2542466394$ & $-0.1723469648$ & $-0.0931417476$ \\
    $\mu^+\mu^-\gamma$ & $p_+$ & $0.3398820615$ & $-0.2994488782$ & $\phantom{-}0.1144873313$ & $-0.0397384416$ \\
                       & $p_-$ & $0.4176202726$ & $\phantom{-}0.3193196432$ & $-0.1095972730$ & $-0.2219603636$ \\
                       & $k$   & $0.2624976659$ & $-0.0198707650$ & $-0.0048900583$ & $\phantom{-}0.2616988052$ \\
    $\pi^+\pi^-\gamma$ & $p_+$ & $0.3413130259$ & $-0.2880589104$ & $\phantom{-}0.1103624989$ & $-0.0430926876$ \\
                       & $p_-$ & $0.4161893082$ & $\phantom{-}0.3079296754$ & $-0.1054724406$ & $-0.2186061176$ \\
                       & $k$   & $0.2624976659$ & $-0.0198707650$ & $-0.0048900583$ & $\phantom{-}0.2616988052$ \\
    \bottomrule
  \end{tabular}
  \caption{Final-state four-momenta in GeV of the phase-space points of \cref{tab:pole}, with
  the $e^+$ beam along $+z$, $E_{e^\pm}=\sqrt s/2$, $\sqrt s=0.7$~GeV for CMD and $1.02$~GeV for
  KLOE small-angle, $p_\pm$ the $\mu^\pm$ or $\pi^\pm$ and $k$ the photon.}
  \label{tab:pole-momenta}
\end{table}

%= bibliography ===================================
\bibliographystyle{JHEP}
\bibliography{journal}
%= end ============================================
\end{document}